\documentclass[a4paper,11pt]{article}
\usepackage{jcappub}
\usepackage[T1]{fontenc} 
\usepackage{orcidlink}

\DeclareUnicodeCharacter{2500}{\textemdash}

\makeatletter
\gdef\@fpheader{Accepted for publication in JCAP}
\makeatother

\usepackage{tcolorbox}
\usepackage{amsmath} 

\arxivnumber{2609.24222} 

\title{\huge{\texttt{StAGE}: Stellar Archaeology-driven Galaxy Evolution II. Binary Black Hole Mergers in Quiescent Galaxies and their Star-forming Progenitors}}

\author[a,b,*]{Irene Iorio\orcidlink{0009-0003-1718-693X},}
\author[a,b,c,d,*]{Andrea Lapi\orcidlink{0000-0002-4882-1735},}
\author[e]{Lumen Boco\orcidlink{0000-0003-3127-922X},}
\author[a,b]{Giovanni Antinozzi\orcidlink{0009-0005-3537-682X},}
\author[a,f]{Michele Bosi\orcidlink{0009-0000-8215-6698},}
\author[e]{Cecilia Sgalletta\orcidlink{0009-0003-7951-4820},}
\author[a,b,g]{Mario Spera\orcidlink{0000-0003-0930-6930},}
\author[a]{Luigi Danese\orcidlink{0000-0003-1186-8430}}

\affiliation[a]{SISSA, Via Bonomea 265, 34136 Trieste, Italy}
\affiliation[b]{INFN-Trieste, Via Valerio 2, 34127 Trieste, Italy}
\affiliation[c]{IFPU, Via Beirut 2, 34014 Trieste, Italy}
\affiliation[d]{INAF-IRA, Via Gobetti 101, 40129 Bologna, Italy}
\affiliation[e]{Univ. Heidelberg, Albert-Ueberle-Str. 3, 69120 Heidelberg, Germany} 
\affiliation[f]{Univ. of Trento, Via Sommarive 14, 38123 Povo (TN), Italy}
\affiliation[g]{INAF-OAR, Via Frascati 33, 00040 Monteporzio Catone (RM), Italy}

\affiliation[*]{corresponding authors}
\emailAdd{iiorio@sissa.it,lapi@sissa.it}

\abstract{We apply \texttt{StAGE}, a data-driven galaxy-evolution framework based on stellar archaeology, to study binary black hole (BBH) mergers associated with present-day quiescent galaxies (QGs) and their star-forming progenitors. By combining \texttt{StAGE} star-formation and chemical-enrichment histories with pre-computed \texttt{SEVN} population-synthesis catalogs, we predict the BBH merger-rate density, intrinsic binary properties, merger-time host-galaxy distributions, and the associated stochastic gravitational-wave (GW) background. A distinctive feature of \texttt{StAGE} is that it jointly tracks the stellar-binary environment at birth and the properties of the host galaxy  at merger time. Moreover, stellar-archaeology constraints on $\alpha$-enhancement allow us to explore how adopting the iron abundance relevant for massive-star winds, rather than oxygen-based total metallicity, affects the predicted BBH merger rates. We find that most BBHs form at moderately subsolar metallicities $Z\lesssim Z_\odot/3$, while their merger-time hosts typically have stellar masses $M_\star\sim10^{10}-10^{11.5}\,M_\odot$ at $z\lesssim2$. At and beyond cosmic noon, hosts lie mainly on the star-forming main sequence, whereas starbursting and quenching descendants contribute increasingly toward lower redshift. Using the independent approach offered by \texttt{StAGE}, we also confirm recent literature findings that, for most binary-evolution prescriptions, predicted local BBH merger rates exceed the LVK-inferred range, while dynamical formation in stellar clusters extends the BBH primary mass distribution into the high-mass regime poorly populated by isolated binaries. Finally, the predicted stochastic background approaches the projected sensitivity of planned LVK upgrades and lies within reach of the Einstein Telescope. All in all, \texttt{StAGE} can provide physically motivated host priors for GW-source localization and dark-siren cosmology.}

\begin{document}
\maketitle
\flushbottom


\defcitealias{Knowles2023}{Kn+23}
\defcitealias{Thomas2010}{Th+10}
\defcitealias{Alvarez2025}{Al+25}

\section{Introduction} \label{sec|intro}

The detection of gravitational waves (GWs) from coalescing compact binaries has opened a new window on the formation and evolution of massive stars, black holes, and galaxies. Since the first binary black hole (BBH) detection \cite{LVKdiscovery}, the LIGO$-$Virgo$-$KAGRA collaboration (LVK; see \cite{LVK2015inst,Virgo2015inst,KAGRA2021inst}) has assembled increasingly larger samples of compact object mergers, enabling statistical constraints on the local merger rate density, the mass and spin distributions of black holes, and the possible redshift evolution of the BBH population \cite{LVK2019_popfirst,LVK2021_popsecond,LVK2023_popthird,LVK2026_popforth}. These observations provide valuable information on the physical processes governing binary stellar evolution \cite{Dvorkin2018,Mapelli2021,Spera2022,Golomb2024,Mould2026}, and on the contribution of different BBH formation pathways \cite{Belczynski2016,Mapelli2020,Zevin2021,Cheng2023,Colloms2025,Sadiq2025,Galaudage2026,Berti2026,Biscoveanu2026}.

A central theoretical challenge is to connect the observed BBH population to its astrophysical birth environment. In this respect, a key quantity is the BBH merger rate density as a function of redshift, of intrinsic compact binary parameters, and of host galaxy properties. In most models, the BBH merger rate is obtained by coupling a metallicity-dependent cosmic star formation history with the outputs of binary evolution calculations. The latter are commonly based either on population synthesis simulations for isolated binaries \cite{Hurley2002,Eldridge2017,Spera2019,Chruslinska2019,Breivik2020,Iorio2023,Andrews2025}, or on semi-analytic/numerical prescriptions for dynamical formation in dense environments \cite{Banerjee2018,DiCarlo2020,Rodriguez2019,Antonini2020,Kremer2020,Torniamenti2024,Kritos2024,Arcasedda2026}.

The cosmic SFR density and its metallicity distribution are usually extracted from hydrodynamical cosmological simulations \cite{Mapelli2017,Lamberts2018,Artale2019,Briel2023,Levina2026}, or inferred from empirical prescriptions based on observed galaxy scaling relations \cite{Dominik2013,Belczynski2016,Boco2019,Neijssel2019,Chruslinska2019,Boco2021,Olejak2021,Santoliquido2022,Broekgaarden2022,Romagnolo2023,Sgalletta2025,Boco2026a, Boco2026b}. These two strategies are somewhat complementary (see review by \cite{Lapi2026SEM} and references therein). Cosmological simulations self-consistently follow the growth of galaxies and their environments, but at the price of substantial computational cost and of subgrid assumptions for processes such as star formation, feedback, chemical enrichment, and stellar evolution. Empirical models are more flexible and computationally efficient, but often compress the galaxy population into global cosmic averages, thereby partially washing out the connection between BBH mergers and their host environments.

The metallicity distribution of the host galaxies is especially important. Stellar winds, remnant masses, pair-instability processes, and the final BBH mass spectrum are all strongly affected by the chemical composition of massive stellar progenitors \cite{Belczynski2016,Spera2017,Dvorkin2016b,Giacobbo2018,Broekgaarden2022,Chruslinska2024}. Moreover, recent studies have pointed out that, at least for metallicities $Z\gtrsim Z_\odot/10$, the most relevant abundance for stellar winds may be closer to that of iron-group elements than to the total metal mass fraction, which is usually inferred from oxygen. This is because iron dominates the line opacity responsible for radiatively-driven mass loss \cite{Vink2001,Vink2022,Chruslinska2025,Romagnolo2026,Boco2026a}. Robust predictions for BBH merger rates therefore require not only a cosmic SFR density, but also a physically motivated description of how star formation is distributed across total metallicity, iron abundance, redshift, and host galaxy properties.

This is a challenging task, and it is therefore not surprising that tensions between theoretical estimates and GW data have progressively emerged. Absolute merger-rate predictions remain somewhat sensitive to the adopted metallicity-dependent star-formation history; nevertheless, several recent studies based on observationally-motivated galaxy evolution recipes concurrently find rates above those inferred by LVK observations \cite{Broekgaarden2022,Santoliquido2022,Srinivasan2023,Romagnolo2023,Boesky2024,Sgalletta2025,Boco2026a,Broekgaarden2026, Boco2026b}.
If confirmed, such a mismatch would point toward the need for revisiting some of the assumptions entering binary stellar evolution, including the onset/treatment of the common envelope phase, stellar winds or natal kicks \cite{Olejak2021,Sgalletta2025,Romagnolo2026,Boco2026a,Boco2026b}. It is therefore important to test this issue with independent, data-driven approaches that minimize assumptions on the galaxy evolution side.

In this work we address the formation and merging of BBHs from an original data-driven perspective. Specifically, we use \texttt{StAGE} (Stellar Archaeology-driven Galaxy Evolution), a semi-empirical framework developed in \cite{Bosi2025} to reconstruct the assembly histories of present-day quiescent galaxies (QGs) and of their star-forming progenitors via stellar archaeology prescriptions. The key idea of \texttt{StAGE} is to exploit the fossil record imprinted in the stellar populations of local QGs $-$ their ages, formation-time distributions, metallicities, and $\alpha$-enhancement $-$ to infer when and how their stars formed. Unlike most semi-empirical models, \texttt{StAGE} does not require an explicit mapping between galaxies and dark-matter halos through abundance matching or related prescriptions. Nor does it require the stellar mass function of QGs to be known at all redshifts. Instead, the redshift-dependent abundance and star formation histories of QG progenitors are reconstructed backward in cosmic time from the statistics of the local QG population and from stellar-archaeology relations \cite{Thomas2005,Thomas2010,Johansson2012,Conroy2014,Knowles2023,Bosi2025}. This makes the framework particularly well suited to applications in which one aims to minimize assumptions on galaxy formation while coupling the resulting histories to independent astrophysical modules, such as stellar and binary population synthesis calculations.

The progenitors of local QGs are especially relevant for BBH formation. Massive QGs dominate a large fraction of the stellar mass budget in the local Universe and host old, $\alpha$-enhanced stellar populations, implying that their stars formed rapidly and predominantly at early cosmic epochs \cite{Thomas2005,Thomas2010,Gallazzi2006, Gallazzi2014, Johansson2012,Conroy2014, vanderWel2014,Moffett2016,Dimauro2019, Driver2022,Davidzon2017,Shuntov2025}. Their progenitors are commonly associated with dusty, compact, intensely star-forming galaxies at cosmic noon and beyond, many of which are faint or invisible in the optical/near-infrared while being bright in the far-infrared and (sub-)millimetre bands \cite{Barro2014,Simpson2014,Talia2021,Behiri2023,Gentile2024stat,Williams2024}. These systems may have contributed substantially to the cosmic SFR density at $z\sim 2-6$ \cite{Casey2018,Zavala2021}, and their high SFRs make them natural birthplaces for massive stellar binaries. Although the descendant galaxies are quiescent today, BBHs formed during their earlier star-forming phase can merge after a broad range of delay times and contribute to the present-day GW event rate.

The first \texttt{StAGE} paper \cite{Bosi2025} showed that the cosmic SFR density inferred for QG progenitors is in remarkable agreement with observational estimates for high-redshift dusty star-forming galaxies, supporting a direct progenitor--descendant connection between these populations. It was also shown that \texttt{StAGE} can reconstruct the average mass and metal assembly histories of QGs in a computationally inexpensive and easily extendable way. Here we use \texttt{StAGE} as the galaxy evolution backbone for BBH formation, and couple its star formation and chemical-enrichment histories to pre-computed binary population-synthesis catalogs built with \texttt{SEVN} \cite{Spera2017,Spera2019,Iorio2023}. The resulting framework provides BBH merger rates and GW observables for the population born in the star-forming progenitors of present-day QGs. Although this population represents only a subset of all galaxies in the Universe, it offers a particularly clean testbed: if the predicted merger rate is already high when only QG progenitors are included, the origin of the mismatch with local LVK data must be sought in other ingredients of the modeling, like in the adopted binary evolution prescriptions.

It is important to stress that the empirical, data-driven inputs entering \texttt{StAGE} necessarily have a finite observational domain of validity. The framework is anchored to the local QG stellar mass function and to stellar-archaeology constraints on ages, age dispersions, and abundance ratios; its high-redshift histories are therefore backward reconstructions from the fossil record rather than direct measurements of the progenitor population at every epoch. Deep galaxy surveys now constrain stellar-mass functions out to $z\gtrsim10$, but number statistics, photometric redshifts, dust corrections, and stellar-mass estimates become progressively less secure toward the earliest epochs \cite{Shuntov2025,Williams2024,Shuntov2026}. Likewise, direct metallicity measurements extend to $z\lesssim 6$ but remain sparse beyond this range, while different strong-line and direct-method calibrations can shift the absolute abundance scale by several tenths of a dex \cite{Kewley2008,Maiolino2019,Curti2023,Nakajima2023}. Consequently, the highest-redshift \texttt{StAGE} predictions should be interpreted as indirectly constrained by the local stellar fossil records and will require direct observational validation as future surveys improve statistics and spectroscopic coverage.

A fundamental advantage of \texttt{StAGE} is that the BBH merger rate can be retained, in a data-driven fashion, as a function of host galaxy properties. Rather than predicting only a total merger rate density, the framework yields distributions in progenitor metallicity, iron abundance, stellar mass, and burstiness, as well as in intrinsic binary parameters such as chirp mass, mass ratio, and component masses. This is relevant for at least two reasons. First, it allows us to identify which regions of the galaxy parameter space dominate the production of BBH mergers, and to assess how sensitive this conclusion is to stellar-archaeology prescriptions and uncertain binary evolution phases such as common envelope. Second, it provides physically motivated host galaxy priors that may be useful for the probabilistic localization of GW sources without electromagnetic counterparts, and hence for dark-siren cosmology \cite{Schutz1986,DelPozzo2012,Fishbach2019,Gray2020,Mastrogiovanni2024}.

In addition to the isolated binary channel, we explore the impact of dynamical evolution in dense stellar environments. Dynamical interactions in star clusters can harden binaries, exchange companions, and promote hierarchical mergers, thereby modifying both the merger efficiency and the mass distribution of BBHs \cite{Rodriguez2016,Rastello2019,Mapelli2020,Antonini2020,Kremer2020,Zevin2021,Weatherford2021,Marchant2024,Oconnor2026,Mai2026}. This contribution is especially relevant at the high-mass end of the primary-BH mass distribution, where isolated binary evolution may underproduce some of the systems inferred from GW observations. We therefore compare the baseline isolated-binary predictions with additional dynamical channels based on the pre-computed BBH catalogs from \cite{Arcasedda2026} generated with the \texttt{BPOP} framework \cite{Arcasedda2020,Arcasedda2023}, to assess to what extent dynamical evolution can populate the high-mass tail.

Finally, the same merger rate model from \texttt{StAGE} can predict the stochastic gravitational-wave background (SGWB) generated by the superposition of unresolved BBH mergers across cosmic time \cite{Phinney2001,Regimbau2011,Rosado2011,Thrane2013,Kowalska2015,Dvorkin2016a,Dvorkin2016b,Perigois2021}. The SGWB is complementary to the resolved merger rate because it is sensitive to the integrated, unresolved BBH population and, in particular, to mergers occurring at high redshift. It can also provide an independent route to cosmological inference \cite{Cousin2026}. The LVK collaboration has only placed upper limits on the SGWB from compact binary mergers \cite{Abbott2018_SGWB}, which are nonetheless informative on the allowed properties of the BBH population \cite{Callister2020,Turbang2024}. Here we compute the SGWB associated with the BBH merger rate predicted by \texttt{StAGE} and compare it with power-law integrated sensitivity curves for present and future detectors, including planned LVK upgrades and the Einstein Telescope (ET; \cite{Branchesi2023}).

The plan of the paper is as follows. In Section~\ref{sec|methods} we describe our basic framework, the construction of the metallicity-dependent SFR density resolved in host galaxy properties, and the computation of BBH merger rates and GW observables. In Section~\ref{sec|results} we present our results on the predicted BBH merger rate density, its dependence on galaxy properties, the intrinsic mass distributions, the impact of the dynamical channel, and the stochastic GW background. In Section~\ref{sec|summary} we discuss and summarize our findings. 

Throughout the paper we assume the standard flat $\Lambda$CDM cosmology from \textit{Planck} \cite{Planck2020}, with Hubble constant $H_0\approx 67.7\,{\rm km\,s^{-1}\,Mpc^{-1}}$, matter density parameter $\Omega_m\approx 0.31$, baryon density $\Omega_b\approx 0.049$ and rms matter fluctuation amplitude $\sigma_8\approx 0.81$ on a scale of $8\, h^{-1}$ Mpc. We also adopt the Kroupa \cite{Kroupa2001} initial mass function (IMF) and the reference solar abundances from \cite{Asplund2021}.

\section{Basic framework} \label{sec|methods}

In this work we employ the semi-empirical framework \texttt{StAGE} \cite{Bosi2025}, originally developed to reconstruct the assembly histories of QGs from stellar archaeological constraints. Here the model is used as the astrophysical backbone for predicting the formation and evolution of BBHs and their GW emission. To this end, the star formation and chemical enrichment histories generated by \texttt{StAGE} are coupled to binary population synthesis calculations, allowing us to follow the evolution of stellar binaries across cosmic time and compute the merger rates of BBHs. For completeness we summarize below the main ingredients of \texttt{StAGE}, referring the reader to \cite{Bosi2025} for a more detailed presentation.

We draw on standard stellar archaeological studies  \cite{Thomas2005,Thomas2010, Gallazzi2006,Gallazzi2014,Conroy2014,Worthley2014,MartinNavarro2018,Morishita2019,Saracco2020,Knowles2023,Beverage2021,Beverage2024,Alvarez2025} to robustly characterize the typical formation time distribution and star formation history of QGs. Specifically, the star formation rate (SFR) for the progenitors of QGs is routinely described by a simple Gaussian parameterization 
\begin{equation}\label{eq|SFRprog}
\psi(z)\equiv \psi(z|M_{\star,\rm QG};t_{\rm form}) = \frac{M_{\star,\rm QG}}{1-R_{\rm ret}}\, \frac{\exp\left[-(t_z-t_{\rm form})^2/2\sigma_{\Delta t}^2\right]}
{\sqrt{2\pi\sigma_{\Delta t}^2}}~,
\end{equation}
so that the corresponding stellar mass $M_\star=(1-R_{\rm ret})\,\int{\rm d}t\,\psi$ reads
\begin{equation}\label{eq|Mstarprog}
M_\star(z)\equiv M_{\star}(z|M_{\star,\rm QG};t_{\rm form}) = \frac{M_{\star,\rm QG}}{2} \left[1+ {\rm erf} \left(
\frac{t_z-t_{\rm form}}{\sqrt{2}\sigma_{\Delta t}}\right) \right]~.
\end{equation}
In the above expressions $M_{\star,\rm QG}$ denotes the relic stellar mass of the galaxy remnant after quenching, $R_{\rm ret}$ is the returned gas fraction from stellar evolution (for a Kroupa IMF in the instantaneous recycling approximation $R_{\rm ret}\approx0.44$ applies), $t_z$ is the cosmic time corresponding to redshift $z$, $t_{\rm form}$ is the formation epoch defined as the cosmic time at which half of the final stellar mass has been accumulated, and $\sigma_{\Delta t}$ characterizes the duration of the main star formation episode. Despite its simplicity, this functional form reproduces the average star formation histories inferred for local QGs over a broad stellar mass range, and provides a convenient and flexible description for generating stellar populations to be evolved with synthesis codes. 

The duration of star formation $\sigma_{\Delta t}$ is constrained by stellar archaeology through the observed $[\alpha/{\rm Fe}]$ abundance. Since $\alpha$ elements are predominantly synthesized by core-collapse supernovae, whereas iron is mainly released by Type-I$a$ supernovae on longer timescales, galaxies experiencing rapid quenching retain enhanced $[\alpha/{\rm Fe}]$ stellar abundances. Stellar archaeological studies therefore provide an empirical calibration between $[\alpha/{\rm Fe}]$ and $\sigma_{\Delta t}$, usually expressed as $[\alpha/{\rm Fe}] \approx \mathcal{C} - \kappa \log \sigma_{\Delta t}\,[{\rm Gyr}],$ where the coefficients $\mathcal{C}$ and $\kappa$ depend only weakly on the adopted stellar yields, Type-I$a$ supernova delay time distribution, IMF, and chemical evolution model \cite{Matteucci1986,Pipino2004,Romano2005,Thomas2005,Thomas2010,Vazdekis2010,Vazdekis2015,DeLucia2014,Vincenzo2016}.

To account for the intrinsic diversity of galaxy assembly histories, in stellar archaeology the formation epoch $t_{\rm form}$ is described through a log-normal probability distribution \cite{Thomas2005,Thomas2010,Johansson2012,Conroy2014,Knowles2023,Alvarez2025}
\begin{equation}
\label{eq|probtform}
\frac{{\rm d}p}{{\rm d}\log t_{\rm form}}(t_{\rm form}|M_{\star,\rm QG}) = \frac{1}{\sqrt{2\pi\sigma_{\log t_{\rm form}}^2}}
\exp\left[-\frac{(\log t_{\rm form}-\log\langle t_{\rm form}\rangle)^2}{2\sigma_{\log t_{\rm form}}^2}\right],
\end{equation}
where $\langle t_{\rm form}\rangle=t_0-\langle t_{\rm age}\rangle$ is set by the average age $\langle t_{\rm age}\rangle$ of the stellar population in the QG and $\sigma_{\log t_{\rm form}}$ is the related age dispersion.

Stellar-archaeology analyses of statistically significant samples of
local QGs provide the average formation epoch $\langle t_{\rm form}\rangle(M_{\star,\rm QG})$, the dispersion $\sigma_{\log t_{\rm form}}(M_{\star,\rm QG})$, and the duration of the star formation $\sigma_{\Delta t}(M_{\star,\rm QG})$, as a function of the relic stellar mass $M_{\star,\rm QG}$ or equivalently of the stellar velocity dispersion $\sigma_\star$ (we can translate between the two quantities using the standard relation by \cite{Cappellari2013}). Throughout this work we consider three representative stellar archaeology calibrations: the classic analysis by \cite{Thomas2010} (hereafter \citetalias{Thomas2010}) based on Lick indices measurements and
models for a sample of individual early-type galaxies from the SDSS; the study by \cite{Knowles2023} (hereafter \citetalias{Knowles2023}) based on Lick indices of stacked SDSS data fitted via a model built with the \texttt{sMILES} stellar library (including
variable $[\alpha/{\rm Fe}]$ ratios); the recent work by \cite{Alvarez2025} (hereafter \citetalias{Alvarez2025}) where the strengths of Lick indices of stacked spectra from the full SDSS Legacy Survey have been fitted via the stellar population models by \cite{Thomas2011}. Throughout the present work, we will present outcomes for these three calibrations in order to assess the sensitivity of our results to current observational and modeling uncertainties in stellar archaeology. 

In the following it will be relevant to consider the star formation history averaged over the formation-time distribution
\begin{equation}\label{eq|psipop}
\langle\psi\rangle(z|M_{\star,\rm QG})= \int {\rm d}\log t_{\rm form}\, \frac{{\rm d}p}{{\rm d}\log t_{\rm form}}(t_{\rm form}|M_{\star,\rm QG}) \, \psi(z|M_{\star,\rm QG};t_{\rm form})~,
\end{equation}
which describes the average evolution of the progenitor
population at fixed relic stellar mass. Note that this should be distinguished from
$\psi(z|M_{\star,\rm QG};\langle t_{\rm form}\rangle)$, which represents
the SFR history of an individual galaxy having the mean formation time
$\langle t_{\rm form}\rangle(M_{\star,\rm QG})$.

\subsection{Cosmic star formation density}
\label{sec|SFRD}

The basic astrophysical input required by binary population synthesis calculations is the cosmic SFR density, together with its distribution in the properties that regulate the formation of compact remnants. Among these, metallicity plays a central role, since it controls stellar winds, remnant masses, and the efficiency with which stellar binaries evolve into merging compact objects. Within the \texttt{StAGE} framework, the contribution from the star-forming progenitors of present-day QGs to the cosmic SFR density is obtained by combining their reconstructed star formation histories with the local stellar mass function of QGs:
\begin{equation}
\label{eq|SFRDqui}
\rho_{\rm SFR}(z)= \int {\rm d}\log M_{\star,\rm QG}\,
\frac{{\rm d}N_{\rm QG}} {{\rm d}\log M_{\star,\rm QG}{\rm d}V}
\int {\rm d}\log t_{\rm form}\, \frac{{\rm d}p}{{\rm d}\log t_{\rm form}}
\, \psi(z)~.
\end{equation}
Here the star formation history $\psi(z)\equiv\psi(z|M_{\star,\rm QG};t_{\rm form})$ is given by Equation (\ref{eq|SFRprog}), and the formation-time distribution follows from Equation (\ref{eq|probtform}). The quantity
${\rm d}N_{\rm QG}/{\rm d}\log M_{\star,\rm QG}\, {\rm d}V$ is the observed stellar mass function of local QGs; we adopt the recent determination by \cite{Shuntov2026} from the COSMOS-Web survey, which is broadly consistent with previous measurements \cite{Moffett2016,Davidzon2017,Driver2022,Weaver2023}.

Since the evolution of massive binaries is strongly metallicity dependent, assigning a realistic chemical abundance to each star-forming progenitor is essential. In keeping with the data-driven philosophy of \texttt{StAGE}, we rely on the observed Fundamental Metallicity Relation (FMR), which links the gas-phase metallicity of star-forming galaxies to their stellar mass and SFR; this relationship has a long tradition in galaxy evolution, and recently it has been tested up to $z\lesssim 6$ by recent JWST observations (see \cite{Mannucci2010,Mannucci2011,Andrews2013,Zahid2014,Hunt2016,Cresci2019,Curti2020,Sanders2021,Chruslinska2021,Curti2023,Nakajima2023,Boco2026a}). In keeping with such data, we describe the metallicity distribution via a log-normal shape
\begin{equation}\label{eq|probzeta}
\frac{{\rm d}p}{{\rm d}\log Z}(Z|\psi,M_\star)=\frac{1}{\sqrt{2\pi\sigma_{\rm FMR}^2}}\exp\left[-\frac{(\log Z-\log Z_{\rm FMR})^2}{2\sigma_{\rm FMR}^2}\right],
\end{equation}
where $\log Z_{\rm FMR}\equiv\log Z_{\rm FMR}(\psi,M_\star)$ is the mean FMR relation and $\sigma_{\rm FMR}\approx0.1$ dex is the  dispersion around it. For the mean FMR we adopt the prescription of \cite{Boco2026a} which, based on the functional form of \cite{Curti2020}, interpolates between the calibration by \cite{Andrews2013} and a progressive saturation at high stellar masses \cite{Chruslinska2021,Boco2026a,Boco2026b}.

The FMR can be regarded as an extension of the more commonly adopted mass--metallicity relation (MZR), with the SFR providing an additional dimension. The two can be connected through the evolving galaxy main sequence, $\psi_{\rm MS}(M_\star,z)$, which specifies the characteristic SFR of galaxies of stellar mass $M_\star$ at redshift $z$ \cite{Daddi2007,Rodighiero2011,Rodighiero2015,Speagle2014,Whitaker2014,Schreiber2015,Mancuso2016b,Dunlop2017,Bisigello2018,Pantoni2019,Lapi2020,Popesso2023,Clarke2024,Rinaldi2025}. In this work we adopt the main-sequence determination by \cite{Popesso2023}, which provides a recent calibration over an extended range of stellar masses and redshifts. Evaluating the FMR along the main sequence, or more generally convolving it with the probability distribution of SFRs around the main sequence, defines an effective redshift-dependent MZR (see \cite{Boco2026a,Boco2026b} for a detailed implementation). At fixed stellar mass, the larger characteristic SFRs of high-redshift galaxies shift them toward the lower-metallicity portion of the FMR. Thus, a substantial part of the observed redshift evolution and scatter of the MZR arises naturally from the combination of the FMR with the evolving distribution of galaxy SFRs. We caution, however, that possible intrinsic evolution of the FMR itself becomes increasingly uncertain at $z\gtrsim6$, while the absolute normalization of both the FMR and MZR remains affected by metallicity-calibration systematics at the level of several tenths of a dex \cite{Mannucci2010,Kewley2008,Curti2020,Maiolino2019,Curti2023,Nakajima2023}.

The metallicity-dependent cosmic SFR density of QG progenitors is written:
\begin{equation}\label{eq|SFRDZprog}
\frac{{\rm d}\rho_{\rm SFR}}{{\rm d}\log Z}(z,Z)=\int{\rm d}\log M_{\star,\rm QG}\frac{{\rm d}N_{\rm QG}}{{\rm d}\log M_{\star,\rm QG}{\rm d}V}\int{\rm d}\log t_{\rm form}\frac{{\rm d}p}{{\rm d}\log t_{\rm form}}\,\psi(z)\,\frac{{\rm d}p}{{\rm d}\log Z}~,
\end{equation}
where the mean metallicity $Z_{\rm FMR}[\psi(z),M_\star(z)]$ is evaluated at the progenitor's SFR $\psi(z)$ and stellar mass $M_\star(z)$. Equation~(\ref{eq|SFRDZprog}) is the main galaxy evolution input supplied to the binary population synthesis calculations described in the following Section.

It is important to distinguish the dispersion $\sigma_{\rm FMR}$ adopted here from the metallicity scatter $\Sigma_{\log Z}$ often used in cosmic-averaged models of the BBH merger rate \cite{Dominik2013,Belczynski2016,Artale2019,Santoliquido2020,Santoliquido2022,Iorio2023} (see also Appendix \ref{app|cosmorate}). These assign an average metallicity $\langle \log Z\rangle(z)$ at each redshift and then introduce a log-normal scatter $\Sigma_{\log Z}$ around it. In those models, $\Sigma_{\log Z}$ describes the full galaxy-to-galaxy metallicity dispersion at fixed redshift. By contrast, $\sigma_{\rm FMR}$ in Equation (\ref{eq|probzeta}) is the residual dispersion around the FMR at fixed stellar mass and SFR. The total metallicity dispersion at a given redshift is therefore not imposed by hand in our framework, but emerges from the integration over the distribution of progenitor stellar masses and SFRs. Plainly, $\Sigma_{\log Z}$ is expected to be substantially larger than $\sigma_{\rm FMR}\approx 0.1$ dex, typically attaining values exceeding $\Sigma_{\log Z}\gtrsim 0.4$ dex. The distinction is relevant to avoid a common misunderstanding, since literature models based on a cosmic-averaged metallicity distribution often adopt an artificially small metallicity scatter $\Sigma_{\log Z}\lesssim 0.15$ dex to reproduce LVK-inferred BBH merger rates, despite the fact this is in clear tension with the observed metallicity distribution \cite{Tang2020,Chruslinska2021,Broekgaarden2022,Boco2026a,Boco2026b}.

Another relevant point concerns the abundance variable entering the BBH merger rate computation. While the total metallicity estimated from oxygen abundance is commonly employed, the formation and evolution of BBHs may correlate more directly with the iron content. In fact, iron-group elements dominate the line opacity of massive star atmospheres and therefore largely regulate radiatively-driven winds; as a result, massive star evolution, BH masses, and BBH merger efficiencies can depend more directly on $Z_{\rm Fe}$ than on the total metal mass fraction $Z$ (see \cite{Chruslinska2024b,Romagnolo2026,Boco2026a}). One possible approach is to empirically calibrate the oxygen-to-iron ratio as a function of specific SFR, used as a proxy for galaxy age (see \cite{Chruslinska2025}). However, such a calibration necessarily combines galaxies with diverse star formation histories and can be affected by substantial scatter and observational systematics.
The \texttt{StAGE} framework provides an alternative route: since the same stellar archaeological constraints that determine the star formation timescale also provide the $\alpha$-enhancement of local QGs, we can infer the iron abundance by combining the total metallicity with the observed $[\alpha/{\rm Fe}]$. To a good approximation, the total metallicity $\log (Z/Z_\odot)$, iron abundance $\log (Z_{\rm Fe}/Z_\odot)\approx [{\rm Fe/H}]$, and $\alpha$-enhancement $[\alpha/{\rm Fe}]$ are related through
\begin{equation}\label{eq|ZFe_map}
\log\left(\frac{Z_{\rm Fe}}{Z_\odot}\right)
= \log\left(\frac{Z}{Z_\odot}\right)-\kappa_Z
\left[\frac{\alpha}{{\rm Fe}}\right]~,
\end{equation}
where $\kappa_Z\approx0.9-1$ depends mildly on the adopted elemental mixture and stellar-population calibration \cite{Salaris1993,Tantalo1998,Trager2000,Thomas2003}.
The iron abundance distribution is approximated by replacing the total-metallicity variable $Z$ with the effective iron-abundance variable $Z_{\rm Fe}$ in Equation~(\ref{eq|SFRDZprog}).

Finally, \texttt{StAGE} can also be exploited to retain information on the properties of the star-forming progenitors. For instance, the cosmic SFR density can be sliced in stellar mass
\begin{equation}\label{eq|SFRDMstar}
\begin{aligned}
\frac{{\rm d}\rho_{\rm SFR}}
{{\rm d}\log M_\star}(z,M_\star) &=
\int{\rm d}\log M_{\star,\rm QG}
\frac{{\rm d}N_{\rm QG}}
{{\rm d}\log M_{\star,\rm QG}{\rm d}V}\times\\
\\
&\times\int {\rm d}\log t_{\rm form}
\frac{{\rm d}p}{{\rm d}\log t_{\rm form}}\,
\psi(z)\, \delta_{\rm D}\left[\log M_\star-\log M_\star(z)\right]~,
\end{aligned}
\end{equation}
where $M_\star(z)\equiv M_\star(z|M_{\star,\rm QG};t_{\rm form})$ is given by Equation (\ref{eq|Mstarprog}).
Hereafter the Dirac delta $\delta_{\rm D}(\log X)$ is numerically implemented as a narrow Gaussian in the variable $\log X$ with a dispersion $\sigma_{\log X}\approx 0.05$ dex.

Beyond stellar mass, it may be useful to characterize the star formation state of the progenitor galaxy. We quantify this in terms of the  offset from the galaxy main sequence $\psi_{\rm MS}(M_\star,z)$, or the burstiness
\begin{equation}\label{eq|Rburst}
\log \mathcal{R}=\log \psi-\log \psi_{\rm MS}(M_\star,z)=\log \left[\frac{\psi}{\psi_{\rm MS}(M_\star,z)}\right]~.
\end{equation}
The corresponding cosmic SFR density sliced in burstiness is
\begin{equation}\label{eq|SFRDR}
\begin{aligned}
\frac{{\rm d}\rho_{\rm SFR}}
{{\rm d}\log \mathcal{R}}(z,\mathcal{R}) &=\int{\rm d}\log M_{\star,\rm QG}\,
\frac{{\rm d}N_{\rm QG}}
{{\rm d}\log M_{\star,\rm QG}{\rm d}V}\times \\
\\
&\times \int{\rm d}\log t_{\rm form}\,
\frac{{\rm d}p}{{\rm d}\log t_{\rm form}}\,\psi(z)\,\delta_{\rm D}\left[\log \mathcal{R}-\log\frac{\psi(z)}{\psi_{\rm MS}[M_\star(z),z]}\right]~.
\end{aligned}
\end{equation}

A similar formalism will be exploited in Section \ref{sec|rates} to derive the BBH merger rate as a function of the host galaxy properties. However, this requires a slightly more involved computation since one needs to take into account the metallicity distribution at the time of BBH formation but pick up the values of the host galaxy properties at the time of the BBH merging (and hence of the GW emission). 

\subsection{Population synthesis simulations for isolated binaries}
\label{sec|SEVN}

The cosmic SFR density described in the previous Section specifies when, where, and at which metallicity stars form. To turn this information into a population of merging BBHs, one needs a model for the evolution of massive stellar binaries. In this Section we focus on the isolated binary evolution channel, while in Section~\ref{sec|BPOP} we investigate how dynamical formation in dense stellar environments affects our baseline results.

We perform binary population synthesis simulations with the code \texttt{SEVN}\footnote{In this work, we use \texttt{SEVN} version 2.17.2. The latest version is publicly available at \href{https://gitlab.com/sevncodes/sevn}{https://gitlab.com/sevncodes/sevn}. Documentation and additional resources are available at \href{https://sevncodes.gitlab.io/sevn}{https://sevncodes.gitlab.io/sevn}.}
(Stellar EVolution for N-body; see \cite{Spera2017,Spera2019,Iorio2023}). 
\texttt{SEVN} evolves single stars by interpolating pre-computed stellar evolution tracks stored in look-up tables, and treats binary interactions through semi-analytic prescriptions. In this work we adopt stellar evolution tracks computed with the latest version of the \texttt{PARSEC} code \cite{Bressan2012,Costa2019,Nguyen2022,Nguyen2025,Costa2025}. Compact remnant formation is modeled using 
the \emph{death matrix} approach (see \cite{Woosley2020}), the \emph{Mapelli20} model for pair-instability and pulsational pair-instability effects (see \cite{Spera2017,Mapelli2020PISN}), and the \emph{unified} natal-kick sampling of \cite{Giacobbo2020}. The stability of mass transfer is assessed by comparing the donor-to-accretor mass ratio with the critical values tabulated by \cite{Hurley2002}.

A major uncertainty in isolated binary evolution is the treatment of the common-envelope (CE) phase, which can strongly affect both the BBH merger efficiency and the delay-time distribution. In \texttt{SEVN}, CE evolution is described through the standard $(\alpha\lambda)$ energy formalism \cite{Webbink1984,Hurley2002}, in which the post-CE separation of the binary is determined by 
\begin{equation}\label{eq|CEalpha}
E_{\rm bind}=\alpha_{\rm CE}\,\Delta E_{\rm orb}~,
\end{equation}
i.e. equating the binding energy of the stellar envelope $E_{\rm bind}$ to the orbital energy $\Delta E_{\rm orb}$ released during the inspiral. The parameter $\alpha_{\rm CE}$ measures the efficiency with which such an orbital energy contributes to envelope ejection. Values $\alpha_{\rm CE}\lesssim 1$ correspond to orbital energy as the dominant reservoir, whereas $\alpha_{\rm CE}>1$ can phenomenologically account for additional energy sources not explicitly included in the baseline formalism \cite{Ivanova2013,Ropke2023}. Because CE evolution remains one of the largest theoretical uncertainties in the formation of compact binaries, we explore several values of the CE parameter, namely $\alpha_{\rm CE}=0.5$, $1$, $3$, and $5$. We also consider a limiting `No CE' case in which CE evolution is suppressed and binaries evolve only through stable mass transfer; note that this involves a separate evolutionary prescription and should not be identified with the formal limit $\alpha_{\rm CE}=0$.

We anticipate that also within \texttt{StAGE}, as in several recent modeling approaches (see Section~\ref{sec|intro}), standard values $\alpha_{\rm CE}\lesssim1$ tend to overpredict the local BBH merger rates inferred by LVK, whereas this tension is substantially alleviated for larger values $\alpha_{\rm CE}\sim 3-5$. We stress, however, that this should not be interpreted as evidence that such high CE efficiencies provide a physically viable solution. Values $\alpha_{\rm CE}\gtrsim 3$ would effectively require very strong additional energy sources contributing to envelope ejection, which are not easy to be physically justified. Moreover, reconstructions of post-CE systems, particularly at low stellar masses, generally favor efficiencies of order unity or below \cite{Ropke2023}. We therefore regard models with $\alpha_{\rm CE}\gtrsim 3$ as extreme, exploratory prescriptions, included here primarily to map the sensitivity of the BBH population to the uncertain treatment of CE evolution.

The initial conditions of the binary population are chosen to describe massive stellar binaries at zero-age main sequence. Primary masses are drawn from the high-mass end of a Kroupa IMF \cite{Kroupa2001},
\begin{equation}
\label{eq|imfprimary}
\frac{{\rm d}p}{{\rm d}m_{\star,1}}
\propto
m_{\star,1}^{-2.3},
\qquad
m_{\star,1}\in[8,150]\,M_\odot~.
\end{equation}
Secondary masses are assigned through the initial mass-ratio distribution measured for massive stars by \cite{Sana2012}:
\begin{equation}
\label{eq|qstarinit}
\frac{{\rm d}p}{{\rm d}q_\star}
\propto
q_\star^{-0.1},
\qquad
q_\star\equiv
\frac{m_{\star,2}}{m_{\star,1}}
\in
\left[
\max\left(\frac{8\,M_\odot}{m_{\star,1}},0.1\right),
1
\right]~.
\end{equation}
The adopted lower bounds in primary mass and mass ratio ensure that both stellar components are drawn from the massive star regime relevant for BBH production. Initial orbital periods are sampled from
\begin{equation}
\label{eq|periodinit}
\frac{{\rm d}p}{{\rm d}\log P_{\rm orb}}
\propto (\log P_{\rm orb})^{-0.55},
\qquad \log\frac{P_{\rm orb}}{{\rm days}}\in[0.15,5.5]~,
\end{equation}
and initial eccentricities from
\begin{equation}
\label{eq|eccinit}
\frac{{\rm d}p}{{\rm d}e} \propto e^{-0.42}, \qquad e\in \left[0,\,1-\left(\frac{P_{\rm orb}}{2\,{\rm days}}\right)^{-2/3}\right]~.
\end{equation}
These distributions follow the observational constraints of \cite{Sana2012}, including the correction discussed by \cite{Moe2017}.

For each CE prescription, we evolve $10^7$ binaries at each of the following 13 metallicities:
$Z=\{0.0002,0.0003,0.0004,0.0007,0.001,0.0014,0.002,0.004,0.007,0.01,0.014,0.02,0.03\}$. The grid is approximately logarithmic, with a somewhat denser sampling of the subsolar regime where most BBH mergers are produced.
For each metallicity, the total simulated stellar mass is
$M_{\star,\rm SIM}\approx 4\times10^8\,M_\odot$. 
The outputs of the population synthesis simulations are conveniently summarized by a merger kernel, which gives the number of BBH merging within a Hubble time produced per unit total stellar mass $M_{\star,\rm SFR}$ formed:
\begin{equation}\label{eq|merg_dist}
\begin{aligned}
\frac{{\rm d}N_{\star\star\rightarrow\bullet\bullet\rightarrow{\bullet}}}{{\rm d}M_{\star,\rm SFR}\,{\rm d}m_{\bullet,1}\,{\rm d}m_{\bullet,2}\,{\rm d}\log\tau}&(m_{\bullet,1},m_{\bullet,2},\tau|Z,\alpha_{\rm CE})=\eta(Z,\alpha_{\rm CE})\times\\
\\
&\times\frac{{\rm d}p}{{\rm d}m_{\bullet,1}\,{\rm d}m_{\bullet,2}\,{\rm d}\log\tau}(m_{\bullet,1},m_{\bullet,2},\tau|Z,\alpha_{\rm CE})~.
\end{aligned}
\end{equation}
Here $m_{\bullet,i}$ are the source-frame component BH masses, with $m_{\bullet,1}\geq m_{\bullet,2}$ by definition, and $\tau$ is the delay time between the formation of the stellar binary and the BBH merger. The probability density ${\rm d}p/{\rm d}m_{\bullet,1}\,{\rm d}m_{\bullet,2}\, {\rm d}\log\tau$ is normalized to unity over the population of systems that merge within the Hubble time, and therefore describes only the shape of the distribution in masses and delay times. The delay time distribution can be obtained a posteriori by integrating over the component masses
\begin{equation}\label{eq|taudist}
\frac{{\rm d}p}{{\rm d}\log \tau} (\tau|Z,\alpha_{\rm CE})= \int{\rm d}m_{\bullet,1}\,{\rm d}m_{\bullet,2}\,\frac{{\rm d}p}{{\rm d}m_{\bullet,1}\,{\rm d}m_{\bullet,2}\,{\rm d}\log\tau}(m_{\bullet,1},m_{\bullet,2},\tau|Z,\alpha_{\rm CE}),
\end{equation}
while integration over delay time and over one of the two mass components yields the corresponding marginal distribution of the other. 

The overall normalization of the merger-kernel is encoded in efficiency $\eta(Z,\alpha_{\rm CE})$, with physical dimension of inverse mass. This quantity is defined as the number of BBH merging within a Hubble time produced per unit total stellar mass formed:
\begin{equation}\label{eq|merg_eff}
\eta(Z,\alpha_{\rm CE})=f_{\rm IMF}\,f_{\rm bin}\,\frac{N_{\bullet\bullet\rightarrow{\bullet},\,\tau<t_{\rm H}}(Z,\alpha_{\rm CE})}{M_{\star,\rm SIM}}~.
\end{equation}
In this expression, $N_{\bullet\bullet\rightarrow{\bullet},\,\tau<t_{\rm H}}$ is the number of simulated binaries that form a BBH and merge within the Hubble time $t_{\rm H}$, while $M_{\star,\rm SIM}$ is the stellar mass explicitly sampled in the simulations. The factor $f_{\rm bin}$ accounts for the assumed binary fraction, and $f_{\rm IMF}\approx M_{\star,\rm SIM}/M_{\star,\rm SFR}$ converts the merger efficiency from the restricted massive star interval sampled by the simulations to the total stellar mass formed according to the adopted IMF. We take $f_{\rm bin}=0.5$ as a conservative choice given the high binary fraction observed among massive stars \cite{Sana2012}, and for our IMF and initial sampling we have $f_{\rm IMF}\approx 0.19$. 

\subsection{BBH merger rate density}\label{sec|rates}

The \texttt{StAGE} framework provides the cosmic SFR density of QG progenitors as a function of redshift and metallicity. We now convolve this astrophysical input with the outputs of the \texttt{SEVN} population synthesis simulations to compute the BBH merger rate density. Specifically, the differential BBH merger rate density at redshift $z$ is obtained as
\begin{equation}\label{eq|Rmerg_components}
\frac{{\rm d} \dot N_{\rm BBH}}{{\rm d}m_{\bullet,1}\,{\rm d}m_{\bullet,2}\,{\rm d}\log Z\,{\rm d}V}=\int{\rm d}\log\tau\,\frac{{\rm d}N_{\star\star\rightarrow\bullet\bullet\rightarrow{\bullet}}}{{\rm d}M_{\star,\rm SFR}\,{\rm d}m_{\bullet,1}\,{\rm d}m_{\bullet,2}\,{\rm d}\log\tau}\,\left[\frac{{\rm d}\rho_{\rm SFR}}{{\rm d}\log Z}\right]_{|t_z-\tau}~.
\end{equation}
Here the SFR density is evaluated at the formation time of the progenitor binary $t_z-\tau$; contributions with $t_z-\tau<0$ are set to zero, equivalently restricting the integral to delay times shorter than the cosmic time available at redshift $z$.

For comparison with GW population studies, it is often convenient to work in terms of the BBH mass ratio $q_\bullet\equiv m_{\bullet,2}/m_{\bullet,1} \leq 1$ and source-frame chirp mass $\mathcal{M}_{\bullet}=(m_{\bullet,1}\,m_{\bullet,2})^{3/5}/(m_{\bullet,1}+m_{\bullet,2})^{1/5}$. A simple change of variables yields
\begin{equation}
\frac{{\rm d}\dot N_{\rm BBH}}
{{\rm d}q_\bullet\,{\rm d}\mathcal{M}_{\bullet}\,{\rm d}\log Z\,{\rm d}V}
=\mathcal{M}_{\bullet}\,
\frac{(1+q_\bullet)^{2/5}}{q_\bullet^{6/5}}
\left[\frac{{\rm d}\dot N_{\rm BBH}}
{{\rm d}m_{\bullet,1}\,{\rm d}m_{\bullet,2}\,{\rm d}\log Z\,{\rm d}V}
\right]_{\left|
\substack{m_{\bullet,1}=\mathcal{M}_\bullet(1+q_\bullet)^{1/5}/q_\bullet^{3/5}\\
\\
m_{\bullet,2}=\mathcal{M}_\bullet(1+q_\bullet)^{1/5}q_\bullet^{2/5}}
\right.}\,.
\end{equation}

Finally, the total BBH merger rate density is obtained by integrating over metallicity and intrinsic binary parameters:
\begin{equation}\label{eq|Rmerg}
\begin{aligned}
\frac{{\rm d}\dot N_{\rm BBH}}{{\rm d}V} & =\int{\rm d}\log Z\int{\rm d}m_{\bullet,1}\int{\rm d}m_{\bullet,2}\,\frac{{\rm d} \dot N_{\rm BBH}}{{\rm d}m_{\bullet,1}\,{\rm d}m_{\bullet,2}\,{\rm d}\log Z\, {\rm d}V} =\\ 
\\
& = \int{\rm d}\log Z\,\eta(Z,\alpha_{\rm CE})\,\int{\rm d}\log \tau\, \frac{{\rm d}p}{{\rm d}\log \tau}(\tau|Z,\alpha_{\rm CE})\,\left[\frac{{\rm d}\rho_{\rm SFR}}{{\rm d}\log Z}\right]_{|t_z-\tau}~,
\end{aligned}
\end{equation}
where in the second equality we have used Equations~(\ref{eq|Rmerg_components}) and~(\ref{eq|taudist}) to make explicit the merger efficiency and the marginalized delay time distribution. Keeping the metallicity dependence instead of integrating over $Z$ gives the BBH merger rate as a function of the progenitor metallicity at binary formation. Similarly, keeping the dependence on $(m_{\bullet,1},m_{\bullet,2})$ or $(q_\bullet,\mathcal{M}_\bullet)$ yields the merger rate distribution in component properties. 
In order to approximately explore the effects of using iron abundances in place of total metallicity (see Section~\ref{sec|SFRD}), we evaluate the pre-computed stellar- and binary-evolution merger kernels by replacing ${\rm d}\rho_{\rm SFR}/{\rm d}\log Z$ with ${\rm d}\rho_{\rm SFR}/{\rm d}\log Z_{\rm Fe}$ in Equation~(\ref{eq|Rmerg_components}). 

A further advantage of \texttt{StAGE} is its ability to retain information on the host galaxy properties of the BBH merger population. This is particularly relevant for host galaxy association, probabilistic localization, and dark-siren cosmology, where the stellar mass or star formation state of the galaxy at the time of BBH merger may provide informative priors. For example, for the stellar mass we can write
\begin{equation}\label{eq|RmergX}
\begin{aligned}
\frac{{\rm d}\dot N_{\rm BBH}}{{\rm d}\log M_\star\,{\rm d}V} &=\int{\rm d}\log M_{\star,\rm QG}\, \frac{{\rm d}N_{\rm QG}}{{\rm d}\log M_{\star,\rm QG}\,{\rm d}V}\, \int{\rm d}\log t_{\rm form}\,\frac{{\rm d}p}{{\rm d}\log t_{\rm form}}\,  \delta_{\rm D}[\log M_\star-\log M_\star(z)]\times\\
&\\
&\times \int{\rm d}\log Z\,\eta(Z,\alpha_{\rm CE})\,\int{\rm d}\log\tau\,\frac{{\rm d}p}{{\rm d}\log\tau}(\tau|Z,\alpha_{\rm CE})\, \left[\psi\,\frac{{\rm d}p}{{\rm d}\log Z}\right]_{|t_z-\tau}
\end{aligned}
\end{equation}
and the same construction can be applied to the burstiness $\mathcal{R}$. Notice that we cannot formally perform the integrations over $M_{\star,\rm QG}$ and $t_{\rm form}$ to reconstruct ${\rm d}\rho_{\rm SFR}/{\rm d}\log Z$ like in Equation (\ref{eq|SFRDZprog}) because the Dirac delta involves the progenitor's stellar mass $M_\star(z)=M_\star(z|M_{\star,\rm QG};t_{\rm form})$ that depends on these quantities. The relevant SFR and metallicity are always those at the binary birth time $t_z-\tau$, because they determine stellar evolution and BBH formation rate. By contrast, the stellar mass or burstiness in Equation~(\ref{eq|RmergX}) are evaluated at the merger time $t_z$, since these are the host galaxy properties relevant for the observed GW events.

\subsection{Impact of dynamical evolution in dense stellar environments}
\label{sec|BPOP}

The isolated-binary channel is not the only possible pathway to BBH formation. Dense stellar systems such as young star clusters (YSCs), globular clusters (GCs) and nuclear star clusters (NSCs) can strongly modify the evolution of compact binaries through repeated dynamical encounters, exchange interactions, binary hardening, ejections, and hierarchical mergers \cite{Mapelli2020,Zevin2021,Cheng2023,Colloms2025,Sadiq2025,Galaudage2026,Berti2026,Biscoveanu2026}. These processes can affect both the normalization of the BBH merger efficiency and the shape of the distributions in component masses, chirp mass, mass ratio, and delay time. In particular, dynamical interactions provide a natural route to populate the high-mass tail of the BBH mass distribution, because massive BHs segregate efficiently toward the cluster centre, interact preferentially, and may undergo repeated mergers if the remnant is retained by the cluster potential.

To estimate the possible impact of dense stellar environments on our results, we use the pre-computed dynamical BBH catalogs from \cite{Arcasedda2026}, generated with the \texttt{BPOP} framework \cite{Arcasedda2020,Arcasedda2023}\footnote{The code is available at \href{https://github.com/marcasedda/BPOP}{{https://github.com/marcasedda/BPOP}}. The catalogs can be found at \href{https://zenodo.org/records/19115567}{https://zenodo.org/records/19115567}.}. \texttt{BPOP} is a semi-analytic population synthesis code designed to model BBH mergers formed  through dynamical interactions within young, globular, and nuclear star clusters. Each cluster is characterized by an initial mass and half-mass radius, from which the core radius, density, velocity dispersion, and escape velocity are derived. The code then follows the long-term structural evolution of the cluster, including mass loss, core contraction and collapse, rebound and expansion, using prescriptions calibrated on direct $N$-body simulations and semi-analytic recipes. The initial cluster masses are sampled from distributions motivated by observed GCs and NSCs, while the YSC mass distribution is modeled as a lower-mass counterpart of the GC distribution; the initial cluster sizes are assigned from observed cluster samples, with a correction accounting for the subsequent secular expansion of clusters. 

Within each evolving cluster, \texttt{BPOP} follows the formation and hardening of BBHs through three-body encounters and binary-single interactions. Dynamical binaries may merge inside the cluster or be ejected and merge later in the field. When a merger occurs, the properties of the remnant BH are computed using numerical-relativity fitting formulae for the final mass, spin, and recoil velocity. If the recoil velocity is smaller than the cluster escape velocity, the merger remnant is retained and can participate in subsequent interactions, leading to higher-generation mergers. The framework also includes the possible formation of very massive BHs from stellar collisions and repeated mergers, which is especially relevant for producing BBHs with primary masses above the range easily accessible to standard isolated binary evolution.

In this work we do not adopt the full cosmic-population model implemented in \texttt{BPOP}, i.e. its prescriptions for the cosmic star-formation
history and for the relative normalization of the different formation environments. Instead, we use the pre-computed \texttt{BPOP} BBH catalogs \cite{Arcasedda2026} to construct merger kernels, in direct analogy with the \texttt{SEVN} kernels introduced above, and convolve them with the \texttt{StAGE} star-formation and metallicity histories of QG progenitors. Thus the cluster dynamics, BBH formation efficiencies, delay time distributions, and mass distributions are taken from \texttt{BPOP}, while the cosmic normalization of the stellar mass formed in YSCs and GCs is specified within the \texttt{StAGE} framework.
For each dense environment we write the dynamical merger kernel as
\begin{equation}\label{eq|merg_dist_dyn}
\frac{{\rm d} N_{\star\star\rightarrow\bullet\bullet\rightarrow\bullet}^{\kappa}}{{\rm d}M_{\star,\rm SFR}^\kappa\,{\rm d}m_{\bullet,1}\,{\rm d}m_{\bullet,2}\,{\rm d}\log\tau}(m_{\bullet,1},m_{\bullet,2},\tau|Z)=\eta_\kappa(Z)\,\frac{{\rm d}p_\kappa}{{\rm d}m_{\bullet,1}\,{\rm d}m_{\bullet,2}\,{\rm d}\log\tau}(m_{\bullet,1},m_{\bullet,2},\tau|Z)~, 
\end{equation}
where $\kappa\in\{{\rm YSC},{\rm GC}\}$, 
$M_{\star,\rm SFR}^\kappa$ is the stellar mass formed in the corresponding clustered environment, $\eta_\kappa(Z)$ is the metallicity-dependent merger efficiency, and the conditional probability density is normalized to unity over the merging BBH population of that channel. All the dependence on the internal cluster population, including the distribution of cluster masses and radii, the cluster structural evolution, binary hardening, ejections, in-cluster mergers, and hierarchical assembly, is already included in the \texttt{BPOP} kernel.

The key additional ingredient required to couple these kernels to \texttt{StAGE} is the amount of stellar mass formed in each clustered environment. For YSCs, we assume that a fixed fraction of the star formation in QG progenitors occurs in young bound clusters. Following the fiducial normalization also adopted in \texttt{BPOP}, we set
\begin{equation}\label{eq|SFRYSC}
\psi_{\rm YSC}(z|M_{\star,\rm QG};t_{\rm form})
= f_{\rm YSC}\,\psi(z|M_{\star,\rm QG};t_{\rm form})~,
\end{equation}
with $f_{\rm YSC}\approx 0.01$ \cite{Bastian2008}. This choice should be interpreted as an effective cluster-formation fraction for the QG progenitor population. Possible dependences on environment and cosmic epoch, which can be expected to some extent, are not modeled explicitly here since they are largely beyond the scope of this paper.

For GCs, we normalize the total stellar mass associated with the cluster population using the empirical relation between the stellar mass of a present-day QG and the total stellar mass in its GC systems \cite{Hudson2014}:
\begin{equation}\label{eq|MGCrel}
\log M_{\star,\rm GC} [M_\odot]
= 7.4+1.6\,(\log M_{\star,\rm QG} [M_\odot]-10.4)~.
\end{equation}
Equivalently, this relation defines a mass-dependent effective GC fraction
\begin{equation}\label{eq|fGC}
f_{\rm GC}(M_{\star,\rm QG})\equiv
\frac{M_{\star,\rm GC}}{M_{\star,\rm QG}}
\simeq2.3\times10^{-3}
\left(\frac{M_{\star,\rm QG}}{10^{11}\,M_\odot}\right)^{0.6}~;
\end{equation}
over the relevant relic-mass interval $M_{\star,\rm QG}\sim10^{10.5}-10^{11.5}\,M_\odot$, this corresponds to $f_{\rm GC}\approx(1.1-4.6)\times10^{-3}$. We then assign to each QG progenitor an effective GC formation history proportional to its reconstructed star formation history
\begin{equation}\label{eq|psiGC}
\psi_{\rm GC}(z|M_{\star,\rm QG};t_{\rm form}) =
f_{\rm GC}(M_{\star,\rm QG})\,
\psi(z|M_{\star,\rm QG};t_{\rm form})~.
\end{equation}
Since the scaling $f_{\rm GC}\propto M_{\star,\rm QG}^{0.6}$ holds, the resulting effective GC formation history is shifted toward earlier epochs relative to the overall \texttt{StAGE} cosmic star-formation history: more massive QGs are assigned larger GC fractions and, through the stellar-archaeology relations, also assemble their stellar populations earlier. The magnitude of this shift depends on the adopted stellar-archaeology prescription. It is modest for \citetalias{Alvarez2025} and \citetalias{Thomas2010}, but more pronounced for \citetalias{Knowles2023},  whose effective GC formation history approaches early-peaked shapes commonly adopted in the literature \cite{Rodriguez2018,Elbadry2019,Mapelli2022}. We stress, however, that also the latter are themselves model dependent, since they rely on assumptions about cluster formation efficiency, galaxy assembly, cluster disruption, and metallicity evolution. In our present implementation, the GC component also inherits the same FMR-based metallicity distribution as the bulk star formation of the corresponding QG progenitor at each formation epoch; namely, we do not impose an additional, preferentially metal-poor GC metallicity distribution.

We clarify that all these assumptions would clearly constitute an oversimplification for a generic galaxy population, since GC formation is generally weighted more strongly toward early, intense star-forming conditions than the overall field-level star formation. For the progenitors of present-day massive QGs, however, the approximation is less severe: both stellar-archaeology constraints and studies of GC systems indicate that the bulk of their stellar mass and a substantial fraction of their GCs formed during the same early, intense assembly phase, although metal-poor GCs may form earlier and/or in lower-mass progenitors subsequently accreted by the final galaxy \cite{Forbes2015,Reina2019,Kruijssen2019}. We therefore regard our treatment as a first-order effective description that captures the broad temporal overlap between rapid QG assembly and GC evolution, rather than as a self-consistent ab-initio model of GC formation.

Then the model construction proceeds exactly as before, in that a metallicity dependent SFR density is derived and then convolved with the merger kernel for the specific dynamical channel from \texttt{BPOP}:
\begin{equation}\label{eq|Rmerg_dyn}
\frac{{\rm d}\dot N_{\rm BBH}^{\kappa}}{{\rm d}m_{\bullet,1}\,{\rm d}m_{\bullet,2}\,{\rm d}\log Z\,{\rm d}V} =\int{\rm d}\log\tau\,\frac{{\rm d}N_{\star\star\rightarrow\bullet\bullet\rightarrow\bullet}^{\kappa}}{{\rm d}M_{\star,\rm SFR}^\kappa\,{\rm d}m_{\bullet,1}\,{\rm d}m_{\bullet,2}\,{\rm d}\log\tau}\, \left[\frac{{\rm d}\rho^\kappa_{\rm SFR}}{{\rm d}\log Z}\right]_{|t_z-\tau}~,
\end{equation}
where again $\kappa\in\{{\rm YSC},{\rm GC}\}$. As in the isolated-binary calculation, the metallicity entering the kernel is the metallicity at the binary formation time, because it determines the stellar-remnant masses and the subsequent dynamical evolution. After integration over metallicity and component masses the total merger rate density including the isolated binaries (IB) and dynamical channels (YSCs, GCs) considered here is therefore:
\begin{equation} \label{eq|Rtot_dyn}
\frac{{\rm d}\dot N_{\rm BBH}^{\rm tot}}{{\rm d} V} = \frac{{\rm d}\dot N_{\rm BBH}^{\rm IB}}{{\rm d} V}+\frac{{\rm d}\dot N_{\rm BBH}^{\rm YSC}}{{\rm d} V}+\frac{{\rm d}\dot N_{\rm BBH}^{\rm GC}}{{\rm d} V}~.
\end{equation}

In the present implementation we do not include NSC channel, because its normalization within the specific population of QG progenitors is more uncertain  \cite{Georgiev2016} and would require additional assumptions about the formation and evolution of NSCs in galaxies of different relic stellar masses (not to mention the presence and destiny of NSCs formed in massive spheroids). 

For this exploratory comparison we fix the common-envelope parameter to the reference value $\alpha_{\rm CE}=1$ and mainly focus on the \citetalias{Thomas2010} stellar-archaeology prescription. This setup isolates the impact of dynamical processing from variations associated with the CE treatment and should primarily be interpreted as a test of how the shapes of the predicted BBH distributions change when dynamical channels are added to the isolated-binary baseline.

\subsection{Stochastic GW background}

The merger rate density derived in the previous Sections describes the intrinsic cosmological population of BBH mergers. For a given detector sensitivity, this population can be separated into individually resolved  and unresolved events. The latter collectively generate a stochastic gravitational-wave background (SGWB).

For each BBH, the matched-filter signal-to-noise ratio (S/N) in a gravitational-wave detector is computed as
\begin{equation}\label{eq|SNR}
{\rm (S/N)}^2= 4\,\int {\rm d}\nu\,
\frac{|\tilde h(\nu)|^2} {S_n(\nu)},
\end{equation}
where $\tilde h(\nu)$ is the Fourier transform of the gravitational waveform and $S_n(\nu)$ is the one-sided detector noise power spectral density. Throughout this work we employ the 
\texttt{IMRPhenomD} inspiral-merger-ringdown waveform model \cite{Husa2016,Khan2016} together with the appropriate detector sensitivity curves for the LVK A\# and ET2026a $10$ km configurations \cite{Gupta2024,Abac2026ET,Dupletsa2026}. A binary is considered individually detected whenever its S/N exceeds a prescribed threshold ${\rm (S/N)}_{\rm thr}$, that for definiteness we take equal to $12$. We define the selection function of resolved events as
\begin{equation}\label{eq|selection}
\mathcal{S}(q_\bullet,\mathcal{M}_\bullet,z) = \Theta_{\rm H}\left[
{\rm (S/N)}(q_\bullet,\mathcal{M}_\bullet,z)-{\rm (S/N)}_{\rm thr}
\right],
\end{equation}
where $\Theta_{\rm H}$ denotes the Heaviside function. The unresolved contribution is then proportional to $1-\mathcal{S}$.

The SGWB is conventionally described by the dimensionless energy-density spectrum
\begin{equation}\label{eq|OmegaGW}
\begin{aligned}
\Omega_{\rm GW}(\nu) & = \frac{\nu}{\rho_cH_0}\, \int{\rm d}z\, \frac{1}{(1+z)E(z)}\times\\
\\
&\times \int {\rm d}q_\bullet\,{\rm d}\mathcal{M}_\bullet\,
\left[1-\mathcal{S}(q_\bullet,\mathcal{M}_\bullet,z)\right]
\frac{{\rm d}\dot N_{\rm BBH}}{{\rm d}q_\bullet\,{\rm d}\mathcal{M_\bullet}\,{\rm d}V}
\,\left[\frac{{\rm d}E_{\rm GW}}{{\rm d}\nu_{\rm s}}\right]_{|\nu_{\rm s}=\nu\,(1+z)},
\end{aligned}
\end{equation}
where $\rho_c=3H_0^2c^2/(8\pi G)$ is the critical density of the Universe, $E(z)\equiv H(z)/H_0$ is the Hubble rate normalized to the present, and ${\rm d}E_{\rm GW}/{\rm d}\nu_{\rm s}$ is the source-frame GW energy spectrum evaluated at the emitted frequency $\nu_{\rm s}=\nu\,(1+z)$. The same expression can be used for the total background by removing the $1-\mathcal{S}$ term inside the integral.

We compare the predicted background with the sensitivity of present and future detector networks through power-law integrated (PI) sensitivity curves \cite{Thrane2013}. These curves are the envelope of all power-law spectra
\begin{equation}
\Omega_\beta(\nu)=\Omega_\beta\left(\frac{\nu}{\nu_{\rm ref}}\right)^\beta
\end{equation}
that would be detected with a prescribed SGWB signal-to-noise ratio ${\rm (S/N)}_{\rm SGWB}$ after an observing time $T$. They can be written as
\begin{equation}
\Omega_{\rm PI}(\nu)=\max_\beta\left[\Omega_{\beta,\rm thr}\left(\frac{\nu}{\nu_{\rm ref}}\right)^\beta\right],
\end{equation}
in terms of the reference frequency $\nu_{\rm ref}=25\,{\rm Hz}$ and of the curve normalization
\begin{equation}\label{eq|SNRSGWB}
\Omega_{\beta,\rm thr}=\cfrac{{\rm (S/N)}_{\rm SGWB}}
{\left[2T\left(\cfrac{3H_0^2}{10\pi^2}\right)^2
\displaystyle\int {\rm d}\nu\,
\cfrac{\gamma^2(\nu)(\nu/\nu_{\rm ref})^{2\beta}}
{\nu^6P_1(\nu)P_2(\nu)}\right]^{1/2}}~.
\end{equation}
Here $\gamma(\nu)$ is the overlap reduction function for a detector pair and $P_1(\nu)$ and $P_2(\nu)$ are the corresponding one-sided noise power spectral densities. In this paper we shall refer to the $1\sigma$ PI sensitivity curves, i.e. ${\rm (S/N)}_{\rm SGWB}=1$, for $T=5$ yr of observations. Note that the PI curves are exploited only to visualize the detectability of the predicted background, but they are not involved in the computation of $\Omega_{\rm GW}$ itself.

\section{Results}\label{sec|results}

In this Section we present the results of our analysis. We first focus on the main ingredients that enter the computation of the BBH merger rate, and then we turn to describe their evolution and dependence on galaxy and compact binary properties, the impact of the dynamical channel on these baseline results, and finally the outcomes on the stochastic GW background.

\begin{figure}[t!]
\centering\includegraphics[width=1.\textwidth]{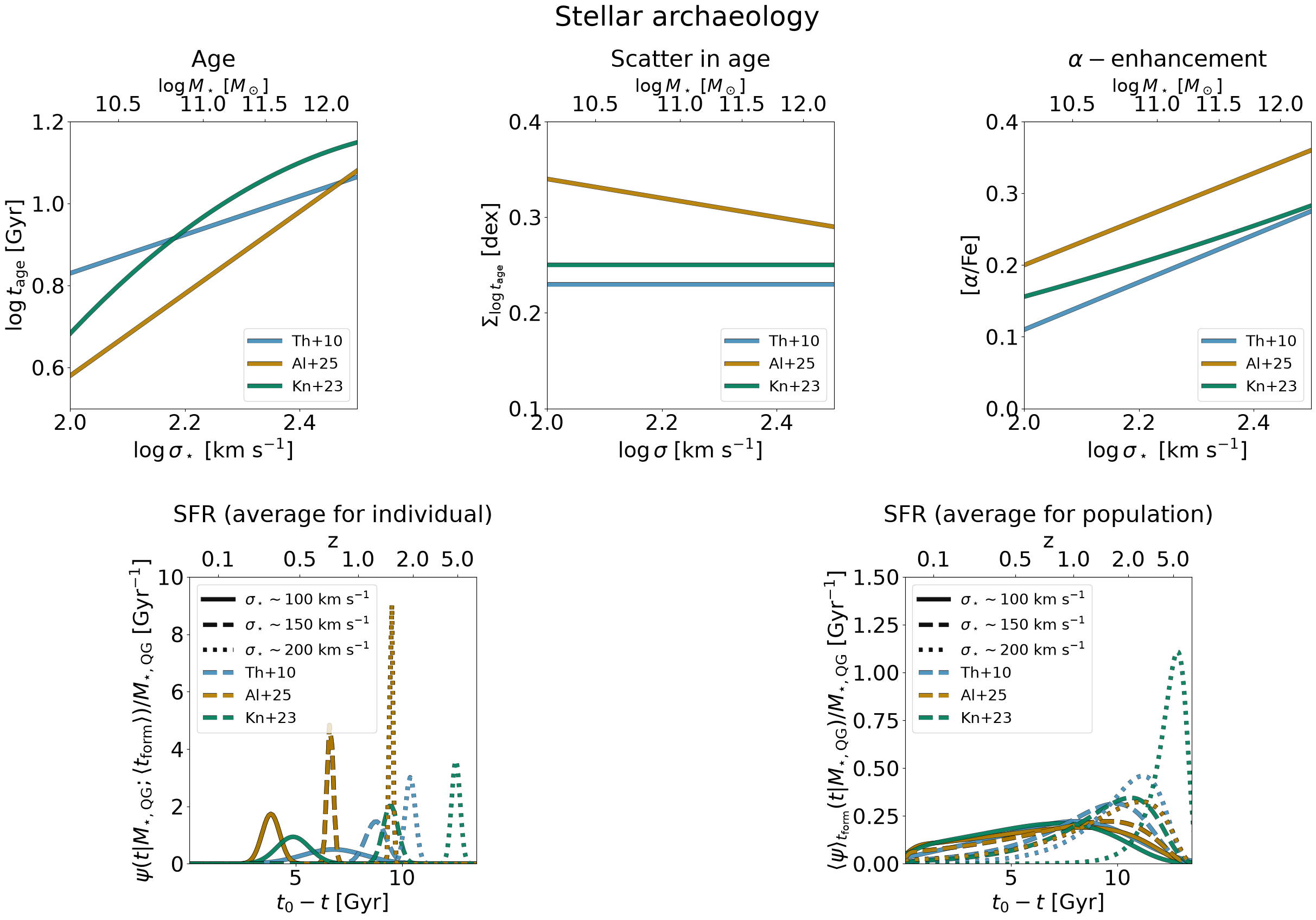}
\caption{Stellar archaeology relationships adopted in \texttt{StAGE}. The top panels illustrate the relations between age (left), scatter in age (middle) and $\alpha-$enhancement (right) vs. the stellar velocity dispersion (bottom $x$ axis) or stellar mass (top $x$ axis). The bottom panels illustrate the corresponding average star formation histories for individual galaxies (left) and for the population (right). Colored lines refer to the stellar archaeology prescription by \citetalias{Thomas2010} (blue), \citetalias{Alvarez2025} (orange) and \citetalias{Knowles2023} (green). In the bottom panels solid, dashed and dotted lines refer to different velocity dispersions $\sigma_\star\approx 100$, $150$, and $200$ km s$^{-1}$, respectively.}\label{fig|SAR}
\end{figure}

\subsection{Basic ingredients}\label{sec|results_basics}

Figure~\ref{fig|SAR} displays and compares the three stellar-archaeology prescriptions adopted in this work. The associated relations between age, age-scatter, and $\alpha$-enhancement vs. stellar velocity dispersion or stellar mass are qualitatively similar in shape but show some quantitative difference in slope and normalization (upper panels). For example, the prescription by \citetalias{Alvarez2025} has an appreciably higher [$\alpha$/Fe] and age dispersion, so implying narrower star formation histories (at given relic mass) than in the other two cases. 
As another example, in terms of age the prescription by \citetalias{Knowles2023} is steeper, which implies that galaxies with high relic masses are substantially older than in the other cases. This behavior is reflected in the star formation histories of individual galaxies (with average formation time), that can differ substantially in their timing and duration (lower left panel). Averaging over the formation-time distribution, however, smooths most of these differences out and produces broadly similar population-averaged histories (lower right panel), although the relative importance of their high-redshift tails remains somewhat prescription dependent.

\begin{figure}[t!]
\centering\includegraphics[width=0.75\textwidth]{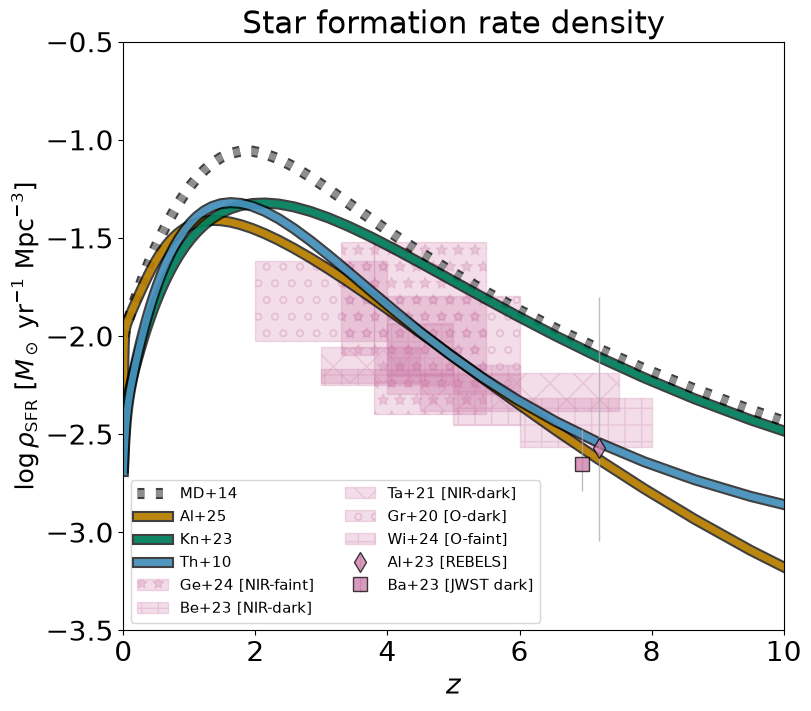}
\caption{The cosmic SFR density as a function of redshift. Solid colored lines show the outcome from \texttt{StAGE} for the progenitors of local quiescent galaxies, when adopting the stellar archaeology prescriptions by \citetalias{Thomas2010} (blue), \citetalias{Alvarez2025} (orange) and \citetalias{Knowles2023} (green). Magenta shaded areas and symbols illustrate the data for dusty star-forming galaxies by \cite{Talia2021,Gruppioni2020,Williams2024,Behiri2023,Gentile2024stat,Barrufet2023,Algera2023}. For reference the dotted grey line is the SFR density for all galaxies by \cite{Madau2014rev}.}\label{fig|SFRD}
\end{figure}

Figure~\ref{fig|SFRD} displays the corresponding cosmic SFR densities as a function of redshift. All three stellar archaeology prescriptions yield a broad maximum around $z\approx 1-2$, followed by a decline toward both lower and higher redshift. The normalization and redshift dependence are consistent with the observational estimates for dust-obscured star-forming galaxies displayed in the figure, while broadly following in shape (with a lower normalization, as expected) the total cosmic SFR density of all galaxy populations (see Appendix \ref{app|cosmorate}). 

Figure~\ref{fig|SFRD_2D_host} further shows that the star formation associated with QG progenitors occupies a relatively well-defined region of galaxy parameter space: the stellar mass distribution is centered at values larger than several $10^{10}\, M_\odot$ in the local Universe, and then progressively smaller toward high redshift; the metallicity distribution is concentrated at moderately subsolar values, with a slightly increasing trend toward low $z\lesssim 2$; most of the star formation density is contributed by galaxies around the main sequence at high redshift, with an appreciable contribution from starbursts and quenched galaxies emerging toward low $z\lesssim 2$. 

\begin{figure}[t!]
\centering\includegraphics[width=\textwidth]{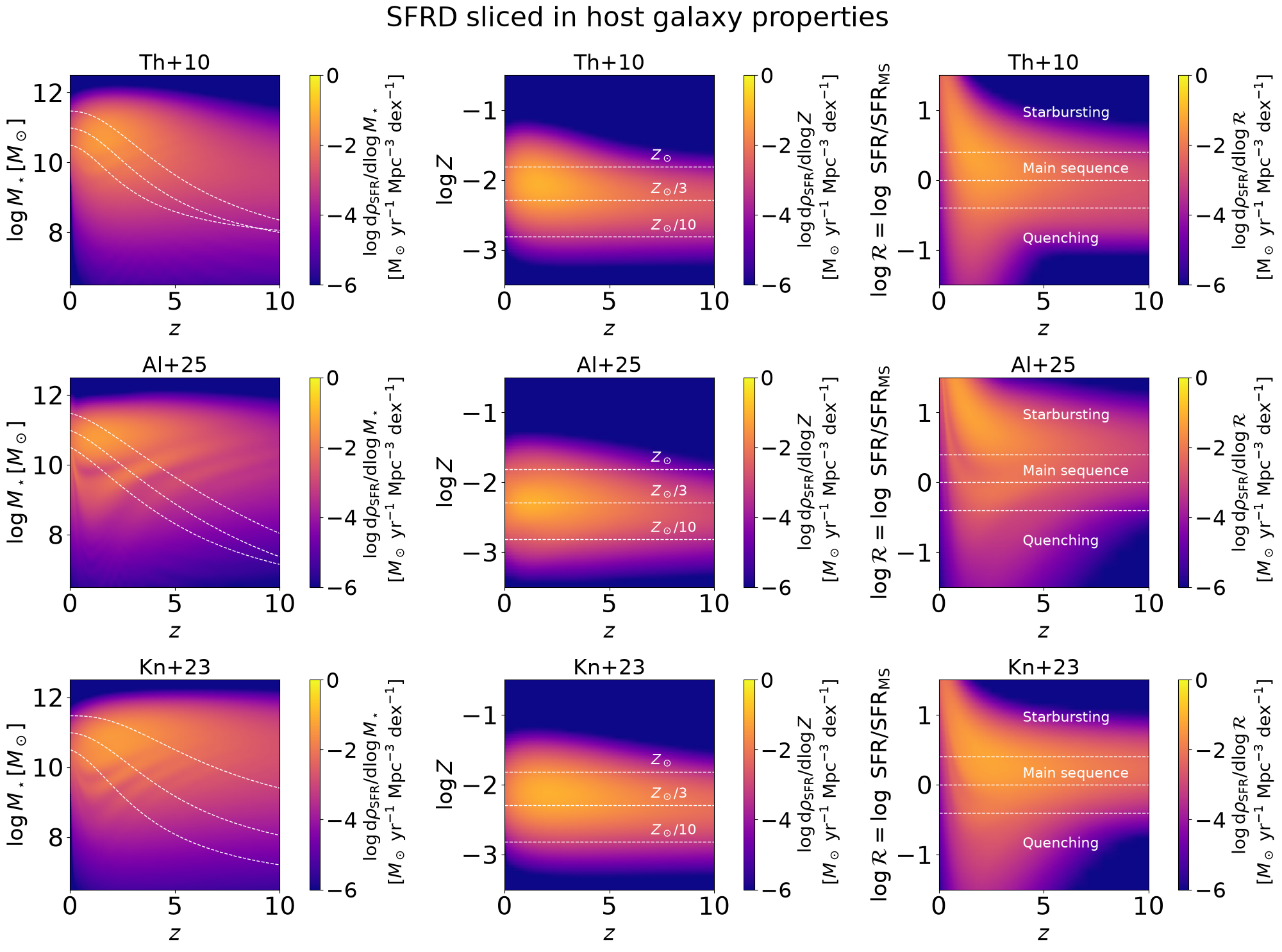}
\caption{The cosmic SFR density from \texttt{StAGE} (color-coded), sliced in stellar mass (left column), metallicity (middle column) and burstiness (right column). Different rows refer to the stellar archaeology prescriptions by \citetalias{Thomas2010} (top), \citetalias{Alvarez2025} (middle) and \citetalias{Knowles2023} (bottom). In the left panels three different average mass growth histories corresponding to final stellar masses of $10^{10.5}-10^{11}-10^{11.5}\, M_\odot$ are also showcased.}\label{fig|SFRD_2D_host}
\end{figure}

We stress that the ridge-like structures visible in the stellar mass and in the burstiness distributions originate from the non-uniform mapping between relic QG properties and the instantaneous masses of their progenitors. At fixed redshift, each combination of $M_{\star,\rm QG}$ and $t_{\rm form}$ is mapped into
$M_\star(z|M_{\star,\rm QG},t_{\rm form})$. Because more massive QGs
assemble earlier and over shorter timescales, different ranges of relic
mass can accumulate around distinct loci in the $(z,M_\star)$ plane,
whereas intermediate stellar masses are traversed comparatively
rapidly and receive less weight. The resulting ridges and intervening
valleys are therefore projection features of the stellar-archaeology
mapping rather than true gaps in the allowed galaxy population. Their prominence depends on the stellar-archaeology prescription and is especially evident for \citetalias{Alvarez2025}, which is characterized by very peaked star formation histories.

\begin{figure}[t!]
\centering\includegraphics[width=0.8\textwidth]{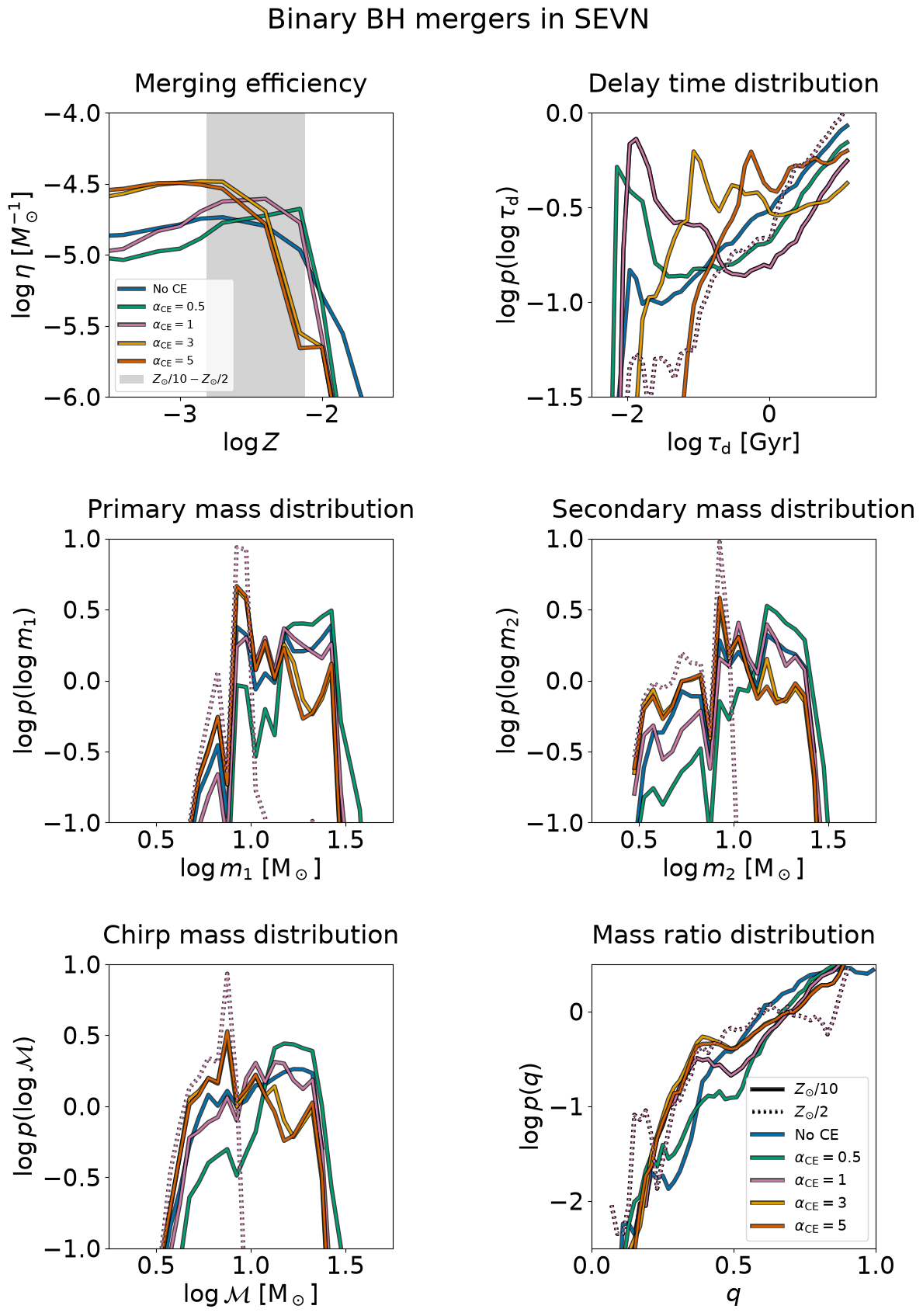}
\caption{Outcomes from the simulations of binary stellar evolution based on the \texttt{SEVN} code: merger efficiency as a function of metallicity (top left), delay time distribution (top right), primary mass distribution (middle left), secondary mass distribution (middle right), chirp mass distribution (bottom left), and mass ratio distribution (bottom right). Colored lines are for different values of the common envelope efficiency parameter: No CE (blue), $\alpha_{\rm CE}=0.5$ (green), $\alpha_{\rm CE}=1$ (magenta), $\alpha_{\rm CE}=3$ (yellow), $\alpha_{\rm CE}=5$ (red). In the top left panel, the grey shaded area illustrates the metallicity range $Z_\odot/10-Z_\odot/2$. In the other panels, solid lines refer to metallicity $Z_{\odot}/10$ and the dotted line to $Z_\odot/2$ (shown only for $\alpha_{\rm CE}=1$).}\label{fig|SEVN}
\end{figure}

Figure~\ref{fig|SEVN} summarizes the second ingredient of the calculation, namely the \texttt{SEVN} merger kernels. The merger efficiency (top left panel) is high at low metallicity and decreases rapidly when $Z\gtrsim Z_\odot/3$. The CE prescription changes both the normalization (especially at the lowest metallicities) and, to a lesser extent, the shape of the efficiency in a non-trivial way. The sizeable efficiency of the No CE model does not imply an absence of binary interactions: stable mass transfer remains active in this case and can provide sufficient orbital hardening for a substantial fraction of systems to merge within a Hubble time.
The delay time distributions (top right panel) are characterized by a double peak: one is at very long delays, and one at relatively small delay times whose location depends on $\alpha_{\rm CE}$ and is present only at low $Z$. The delay time distributions are structured and cannot in general be reduced to a single featureless ${\rm d}p/{\rm d}\tau\propto\tau^{-1}$ law. The component-mass and chirp-mass distributions (middle and lower panels) are also strongly metallicity dependent, but even at low $Z\lesssim Z_\odot/10$ the most massive remnants do not exceed $40-50\, M_\odot$. All models independently of $\alpha_{\rm CE}$ and metallicity favor relatively large mass ratios. These trends determine how the star formation shown in Figure~\ref{fig|SFRD} is mapped into the BBH merger rate distributions discussed below.

\begin{figure}[t!]
\centering\includegraphics[width=0.725\textwidth]{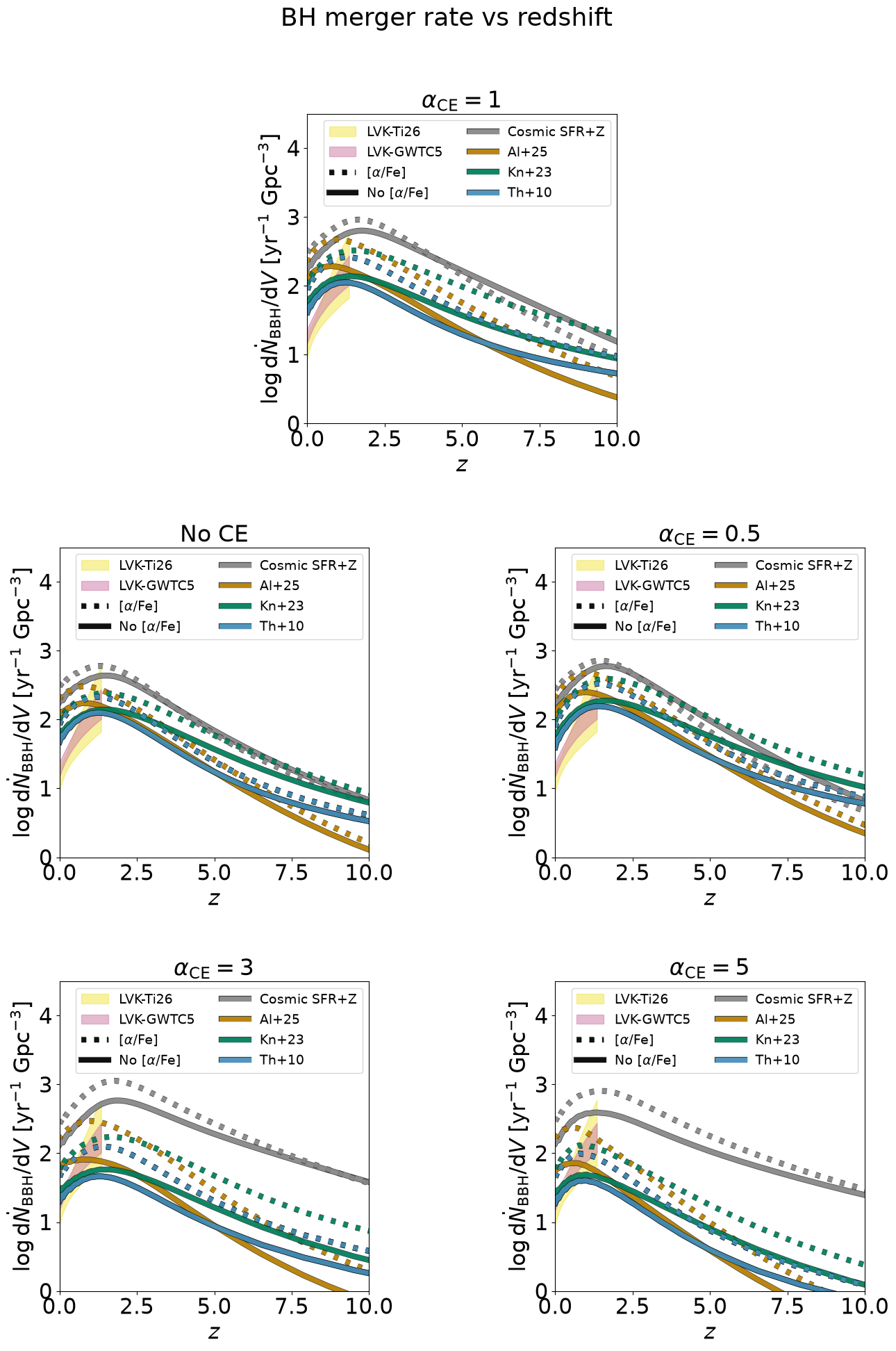}
\caption{The cosmic merger rate density of BBHs as a function of redshift. Various panels refer to different values of the common envelope efficiency parameter $\alpha_{\rm CE} = 1$ (reference value, top), No CE (middle left), $\alpha_{\rm CE} = 0.5$ (middle right), $\alpha_{\rm CE} = 3$ (bottom left), $\alpha_{\rm CE} = 5$ (bottom, right). Colored lines show the outcomes from \texttt{StAGE} for the progenitors of massive quiescent galaxies, when adopting the stellar archaeology prescription by \citetalias{Thomas2010} (blue), \citetalias{Alvarez2025} (orange) and \citetalias{Knowles2023} (green). Solid lines use the oxygen-based metallicity $Z$, whereas dotted lines use the $\alpha$-corrected iron abundance $Z_{\rm Fe}$. Data by LVK are from GWTC-5.0 official catalog (\textit{Default-BBH} in \cite{LVK2026_popforth}; magenta shaded area) and by the \texttt{Vamana} mixture-model (\cite{Tiwari2026}; yellow shaded area). For reference the grey lines refer to the total merger rate obtained by combining the cosmic SFR density for all galaxies with an average cosmic metallicity evolution (see Appendix \ref{app|cosmorate} for details).}\label{fig|BBHMR}
\end{figure}

\subsection{Cosmic BBH merger rate and host galaxy properties}
\label{sec|results_rates}

Figure~\ref{fig|BBHMR} presents the cosmic BBH merger rate density as a function of redshift. For all stellar-archaeology prescriptions, the rate rises from the local Universe to a broad maximum at $z\approx1-2$ and subsequently decreases toward higher redshift. The maximum occurs later than the main star formation episode of the QG progenitors because the merger history is broadened and shifted by the BBH delay time distribution. Nevertheless, the rate retains an extended high-redshift tail, especially for the prescriptions with a larger amount of early star formation.

For the reference value $\alpha_{\rm CE}=1$, as well as for the No CE and $\alpha_{\rm CE}=0.5$ models, the local merger-rate density produced by the progenitors of QGs alone lies substantially above the ranges inferred from LVK data. This conclusion is robust against the adopted stellar-archaeology calibration: the three prescriptions introduce variations of at most a factor of a few, substantially smaller than the discrepancy with the observed local normalization. Increasing the CE efficiency to $\alpha_{\rm CE}=3-5$ lowers the predicted merger rate and brings it appreciably closer to the LVK intervals, although the precise level of agreement still depends on the stellar-archaeology and abundance prescriptions. As already mentioned, such an improved agreement with the LVK rates should not be taken as an evidence for the physical viability of high values $\alpha_{\rm CE}\gtrsim 3$, which are rather extreme and are reported only as exploratory prescriptions to map the sensitivity of the BBH population to the uncertain treatment of CE evolution.

In addition, the dependence on $\alpha_{\rm CE}$ is intrinsically non-trivial, because the CE prescription affects both the probability that a binary survives the envelope phase and the orbital separation of the surviving system. In our models, the lower rates obtained for $\alpha_{\rm CE}=3-5$ result from two related features of the merger kernels. First, over the metallicity range $Z\sim Z_\odot/10-Z_\odot/3$ that dominates BBH production in QG progenitors, the merger efficiency declines more steeply with increasing $Z$ for the larger values of $\alpha_{\rm CE}$. Second, increasing $\alpha_{\rm CE}$ shifts a substantial fraction of the delay time distribution toward longer $\tau_{\rm d}$, as expected because more efficient envelope ejection generally leaves surviving binaries at wider post-CE separations. The resulting merger rate therefore reflects a non-trivial convolution between the metallicity-dependent efficiency, the delay time distribution, and the cosmic SFR density at the corresponding formation epoch. 
In the present models, this interplay ultimately produces the somewhat lower BBH merger rates found for the largest values of $\alpha_{\rm CE}$. 

The dotted curves in Figure~\ref{fig|BBHMR} illustrate the effect of interpreting the metallicity relevant for massive star evolution in terms of the $\alpha$-corrected iron abundance. Within our effective-$Z_{\rm Fe}$ approximation, an $\alpha$-enhanced population is associated with a lower iron abundance at fixed total metallicity. Evaluating the merger kernels at this lower effective abundance increases the inferred BBH formation efficiency and hence the predicted merger rate. Thus, accounting for $\alpha$-enhancement strengthens the rate excess with respect to the data. This comparison illustrates that the predicted BBH merger rate can be sensitive to whether the relevant abundance scale is associated with total metallicity or with iron content in rapidly formed, $\alpha$-enhanced stellar populations.

The impact of the $\alpha$-enhancement correction is somewhat stronger for the largest values of $\alpha_{\rm CE}$ because the corresponding \texttt{SEVN} merger efficiency has a steeper dependence on metallicity over the range $Z\sim Z_\odot/10-Z_\odot/3$ that dominates BBH formation in our model (see Figure~\ref{fig|SEVN}). At fixed total metallicity, replacing $Z$ with the lower effective iron abundance $Z_{\rm Fe}$ therefore produces a larger increase in the merger efficiency for $\alpha_{\rm CE}\sim 3-5$ than for the lower-$\alpha_{\rm CE}$ prescriptions. This effect is then propagated through the convolution with the metallicity-dependent star-formation history into the merger rates shown in Figure~\ref{fig|BBHMR}.

For reference, the grey curves in Figure~\ref{fig|BBHMR} show the rate obtained by combining the SFR density of all galaxies with an average cosmic metallicity evolution (see Appendix \ref{app|cosmorate} for details). As expected, these curves lie above the predictions based only on QG progenitors. However, the difference is modest compared with the large reduction in the galaxy population included in \texttt{StAGE}. The star-forming progenitors of present-day QGs can therefore account for a substantial fraction of the total BBH merger budget. The fact that this restricted population already tends to overproduce the observed local rate makes it difficult to attribute the tension solely to uncertainties in the global cosmic star formation history. These findings therefore suggest that some of the basic assumptions entering stellar and binary evolution $-$ for example, CE onset/evolution, stellar winds, natal kicks, etc. $-$ may need to be reconsidered. In this respect, our results are consistent with recent studies based on different approaches that have also found merger-rate normalizations above the LVK-inferred range for several population-synthesis assumptions \cite{Broekgaarden2022,Santoliquido2022,Srinivasan2023,Boesky2024,Sgalletta2025,Boco2026a,Boco2026b}.  

\begin{figure}[t!]
\centering\includegraphics[width=0.75\textwidth]{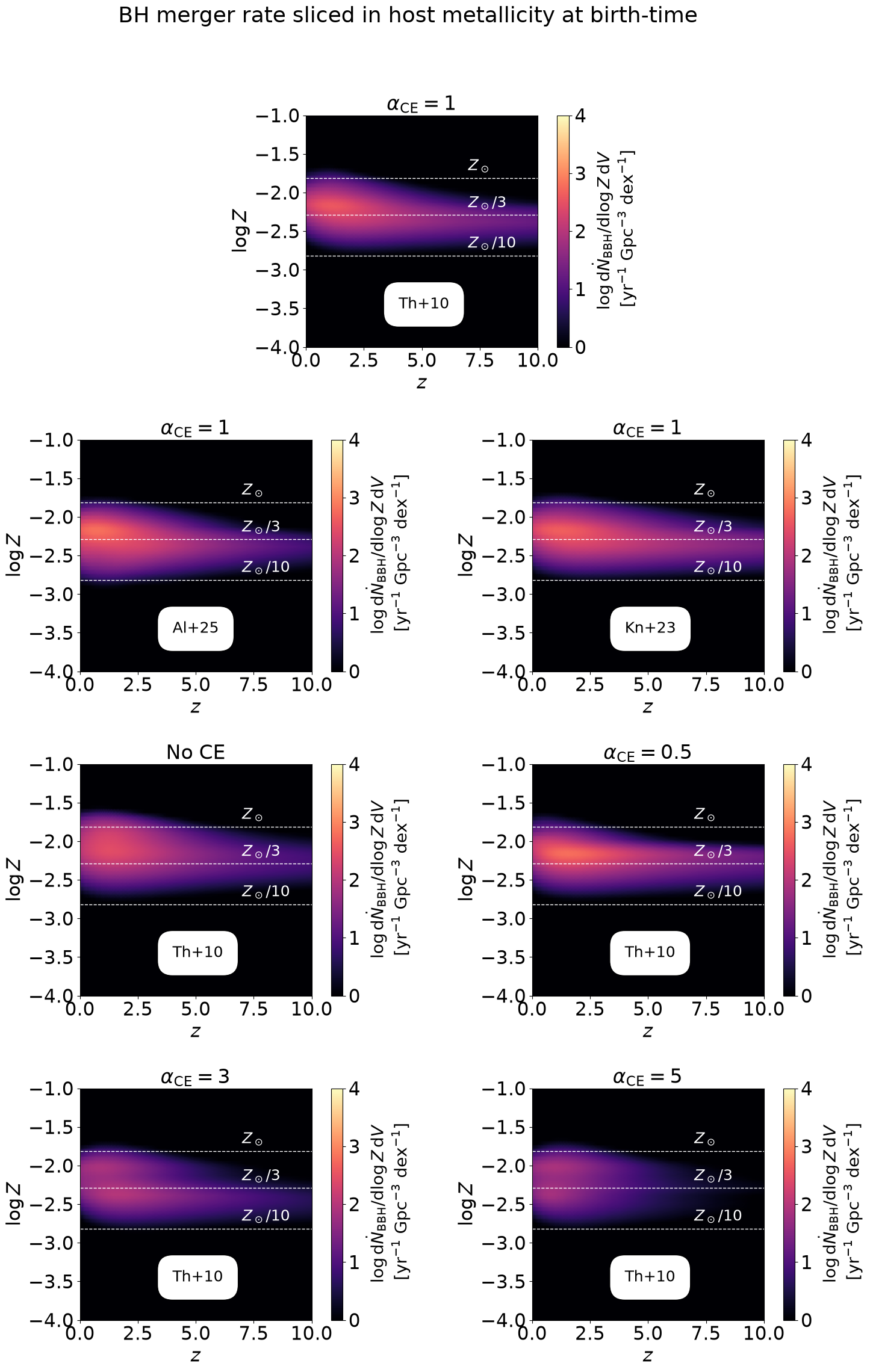}
\caption{The cosmic merger rate density of BBHs (color coded) sliced in metallicity at the birth time and redshift. The top single panel is for a reference combination of the stellar archaeological prescription by \citetalias{Thomas2010} and common envelope parameter $\alpha_{\rm CE}=1$. The second row of panels varies the prescriptions to that of \citetalias{Alvarez2025} (left) and \citetalias{Knowles2023} (right), while keeping $\alpha_{\rm CE}=1$. The two bottom rows of panels show the No CE and $\alpha_{\rm CE}=0.5$, $3$ and $5$ cases, while keeping the prescription by \citetalias{Thomas2010}.}\label{fig|BBHMR_2D_Z}
\end{figure}

\begin{figure}[t!]
\centering\includegraphics[width=0.8\textwidth]{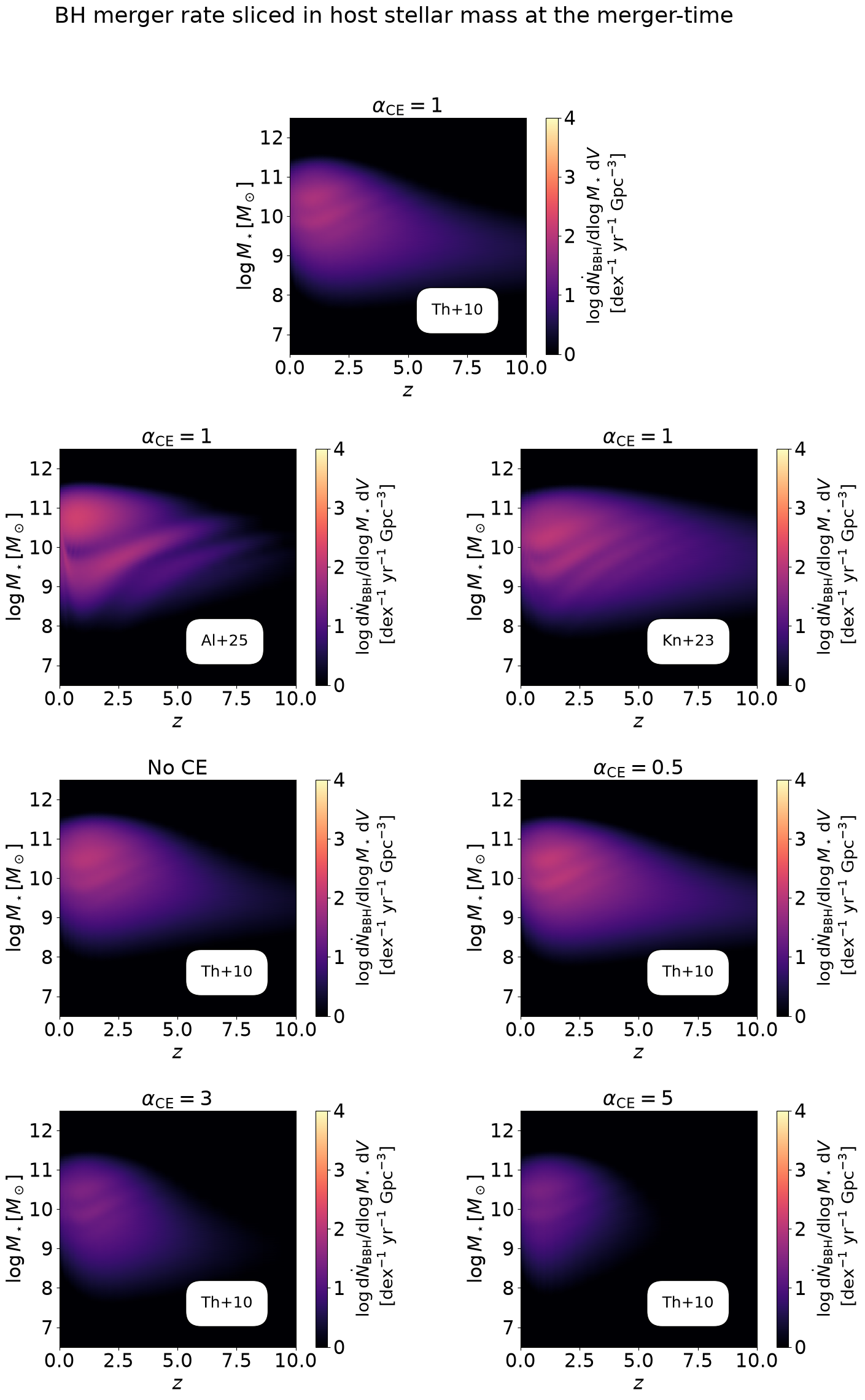}
\caption{Cosmic merger rate density of BBHs sliced in stellar mass at the merger time and redshift. Panels as in Figure \ref{fig|BBHMR_2D_Z}.}\label{fig|BBHMR_2D_Mstar}
\end{figure}

\begin{figure}[t!]
\centering\includegraphics[width=0.825\textwidth]{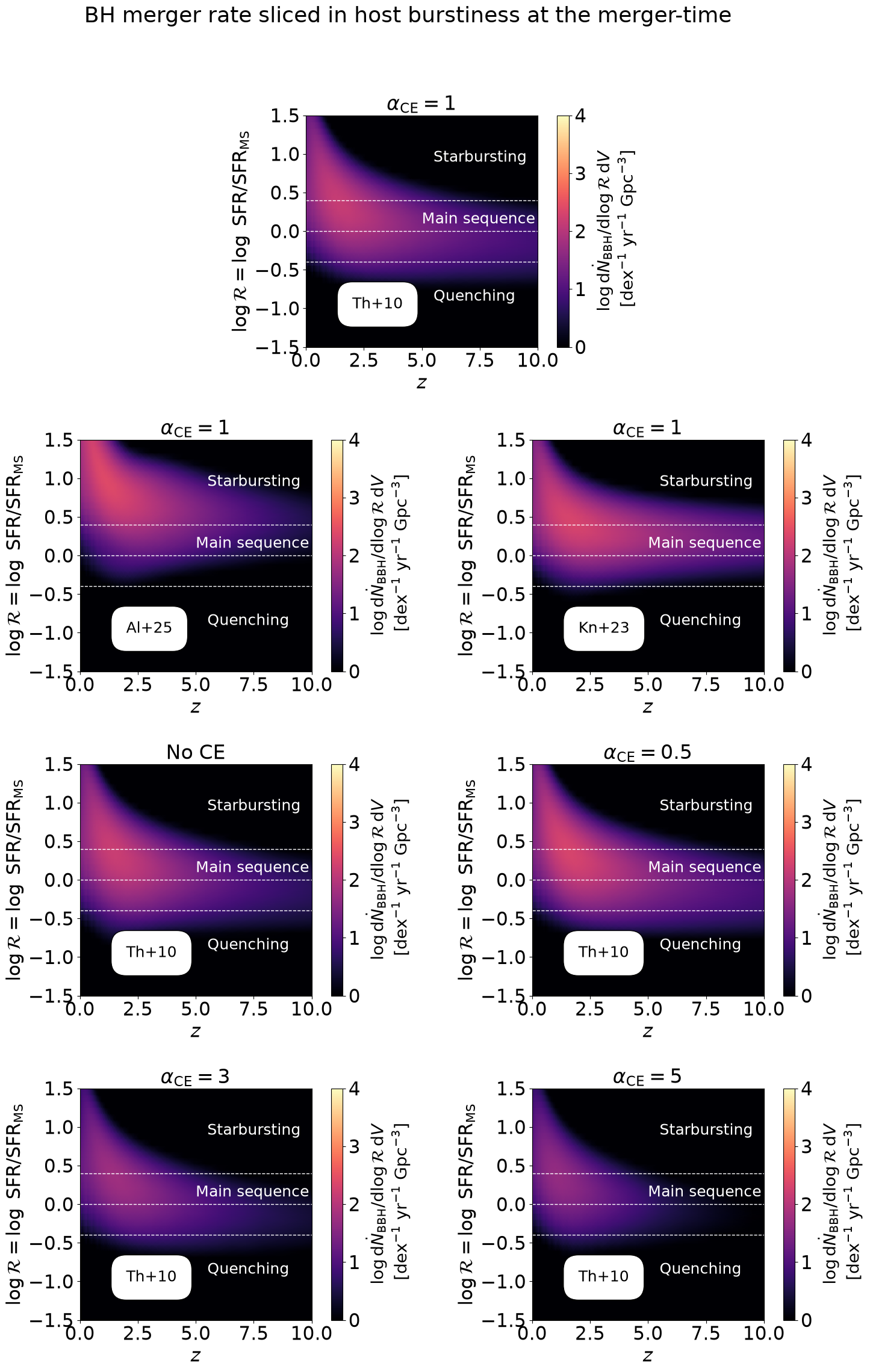}
\caption{Cosmic merger rate density of BBHs sliced in burstiness at the merger time and redshift. Panels as in Figure \ref{fig|BBHMR_2D_Z}.}\label{fig|BBHMR_2D_Rburst}
\end{figure}

Figure~\ref{fig|BBHMR_2D_Z} resolves the merger rate density in the metallicity of the host at the binary birth time. The dominant contribution is concentrated between approximately $Z_\odot/10$ and $Z_\odot/3$, with a tail toward $Z_\odot/2$ at low redshift. The locus shifts gradually toward lower metallicity with increasing redshift. This behavior results from the competition between the increasing efficiency of BBH production toward low $Z$ and the declining amount of metal-poor star formation in the progenitors of massive QGs. The location of the dominant metallicity band is only weakly affected by $\alpha_{\rm CE}$, which mainly changes its normalization, whereas the stellar-archaeology prescription affects the relative strength and extent of the high-$z$ contribution.

The stellar mass and burstiness distributions shown in Figures~\ref{fig|BBHMR_2D_Mstar} and~\ref{fig|BBHMR_2D_Rburst} refer instead to the galaxy at the merger epoch.  Figure~\ref{fig|BBHMR_2D_Mstar} shows that mergers at $z\lesssim2$ occur predominantly in descendants with stellar masses $M_\star\sim10^{10}-10^{11.5}\,M_\odot$; at progressively higher redshift the dominant locus moves toward $M_\star\sim10^{9}-10^{10.5}\,M_\odot$, because the progenitors have not yet assembled their final stellar mass.  Changing $\alpha_{\rm CE}$ primarily rescales the rate and alters its redshift reach, while leaving the characteristic host-mass sequence broadly unchanged.

The burstiness distribution in Figure~\ref{fig|BBHMR_2D_Rburst} provides complementary information on the evolutionary state of the merger host. Around redshifts $z\sim 1-3$ at which the merger rate is largest, the dominant hosts lie on the star-forming main sequence or in the starbursting region. Toward higher redshift the main locus approaches the main sequence, while toward the local Universe an appreciable tail extends into the quenching regime. The latter is the natural signature of long-delay binaries: they formed when the progenitor was actively assembling stars but merge only after the descendant has moved below the main sequence and become quiescent. Hence, the same BBH population can be associated with a star-forming birth environment and a quiescent host at coalescence. 

These host-property distributions constitute one of the distinctive outputs of the data-driven approach underlying \texttt{StAGE}. The birth metallicity identifies the conditions that regulate stellar and binary evolution, whereas the stellar mass and burstiness at merger time describe the galaxy relevant for electromagnetic follow-up, statistical host association, and dark-siren analyses. For BBHs associated with the progenitors of present-day QGs, these host properties evolve appreciably with redshift: low-redshift mergers occur predominantly in massive descendants spanning a broad range of star-formation states, from ongoing starbursts to quenching systems depending on the adopted stellar-archaeology prescription, whereas toward higher redshift the hosts shift to lower stellar masses and increasingly approach the star-forming main sequence.

In more detail, it is worth to quantify the starburst fraction of merger-time hosts. Defining galaxies as starbursting when $\log\mathcal{R}>\log \mathcal{R}_{\rm SB}\approx 0.4$ dex, the starburst fraction can be computed as
\begin{equation}\label{eq|fSB}
f_{\rm SB}(z)=
\frac{\displaystyle\int_{\log \mathcal{R}_{\rm SB}}^{+\infty}{\rm d}\log\mathcal{R}\,
\frac{{\rm d}^2\dot N_{\rm BBH}}{{\rm d}V\,{\rm d}\log\mathcal{R}}}
{\displaystyle\int_{-\infty}^{+\infty}{\rm d}\log\mathcal{R}\,
\frac{{\rm d}^2\dot N_{\rm BBH}}{{\rm d}V\,{\rm d}\log\mathcal{R}}}~.
\end{equation}
The corresponding fraction depends considerably on the adopted stellar archaeology prescription. For the reference $\alpha_{\rm CE}=1$ case, at $z\approx \{0.2,1.5,3,6\}$, we find respectively $f_{\rm SB}\approx \{0.7,0.3,0.1,0.03\}$ for \citetalias{Thomas2010}, $\{0.97,0.9,0.7,0.6\}$ for \citetalias{Alvarez2025}, and $\{0.8,0.5,0.3,0.2\}$ for \citetalias{Knowles2023}. The particularly large fractions obtained for \citetalias{Alvarez2025} reflect the very short star-formation timescales inferred by this prescription: galaxies undergo an intense phase lying well above the main sequence and subsequently quench rapidly, spending comparatively little time in an intermediate star-forming state. By contrast, the dependence of these fractions on $\alpha_{\rm CE}$ is weak. This illustrates that, while binary-evolution prescriptions strongly affect the overall merger-rate normalization, the predicted distribution among different host star-formation states is much more sensitive to the adopted galaxy evolution prescriptions. More generally, variations in the galaxy sector can partially mimic variations in stellar and binary evolution, producing a degeneracy in both merger rates and host-galaxy properties \cite{Broekgaarden2022,Srinivasan2023,Broekgaarden2026}. The stellar-archaeology anchoring of \texttt{StAGE} reduces the freedom on the galaxy-evolution side for the QG-progenitor population, although uncertainties associated with high-redshift star formation and metallicity are not totally eliminated. In this respect, the stochastic GW background provides a complementary constraint because it retains sensitivity to the numerous unresolved mergers at higher redshift, where alternative star-formation and delay-time prescriptions can differ most strongly \cite{Dvorkin2016a,Dvorkin2016b,Callister2020,Turbang2024}.

\begin{figure}[t!]
\centering\includegraphics[width=0.725\textwidth]{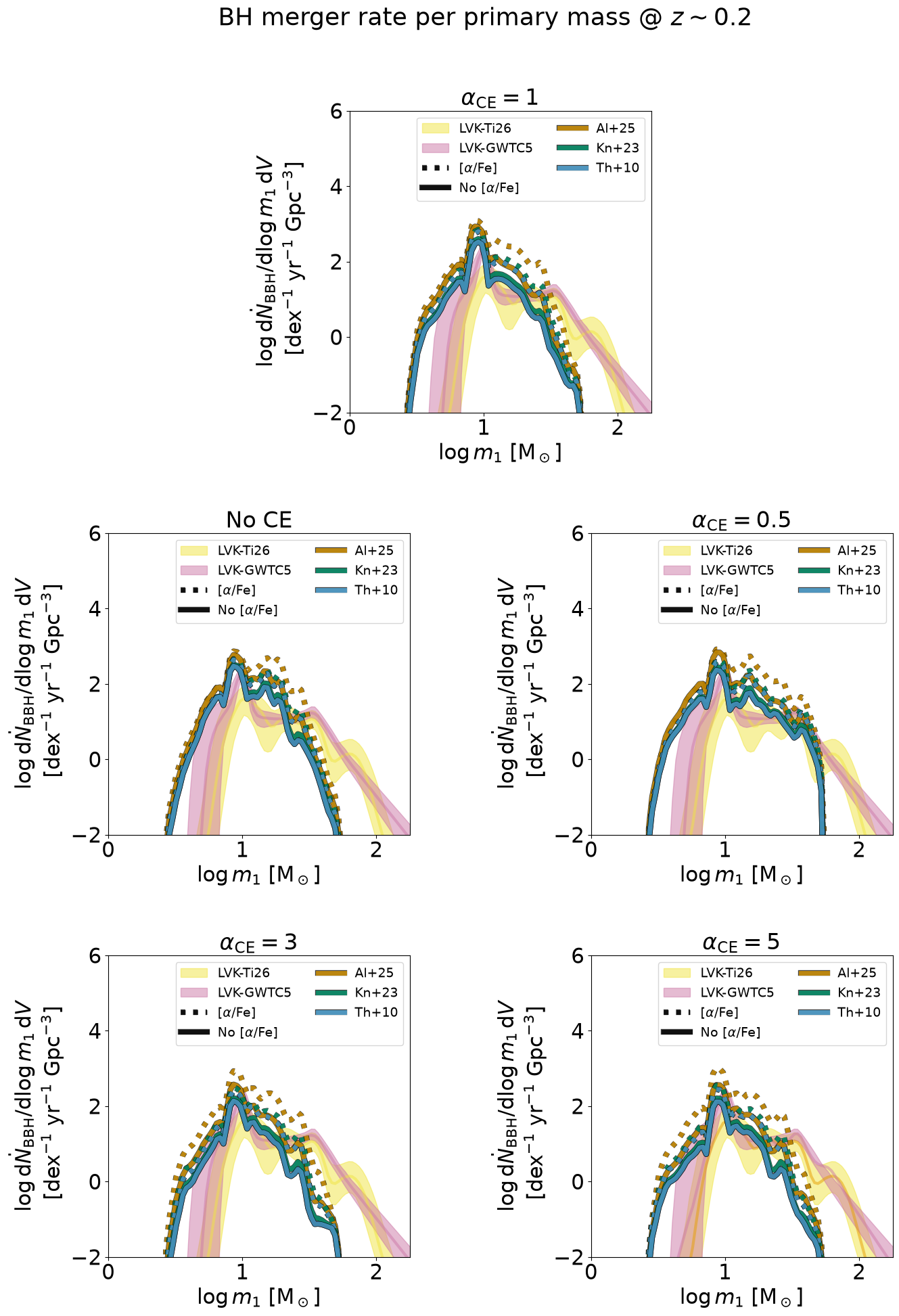}
\caption{Primary mass distribution of BBH mergers at $z\approx 0.2$. Panels, colors and linestyles as in  Figure \ref{fig|BBHMR}.}\label{fig|BBHMR_m1}
\end{figure}

\begin{figure}[t!]
\centering\includegraphics[width=0.725\textwidth]{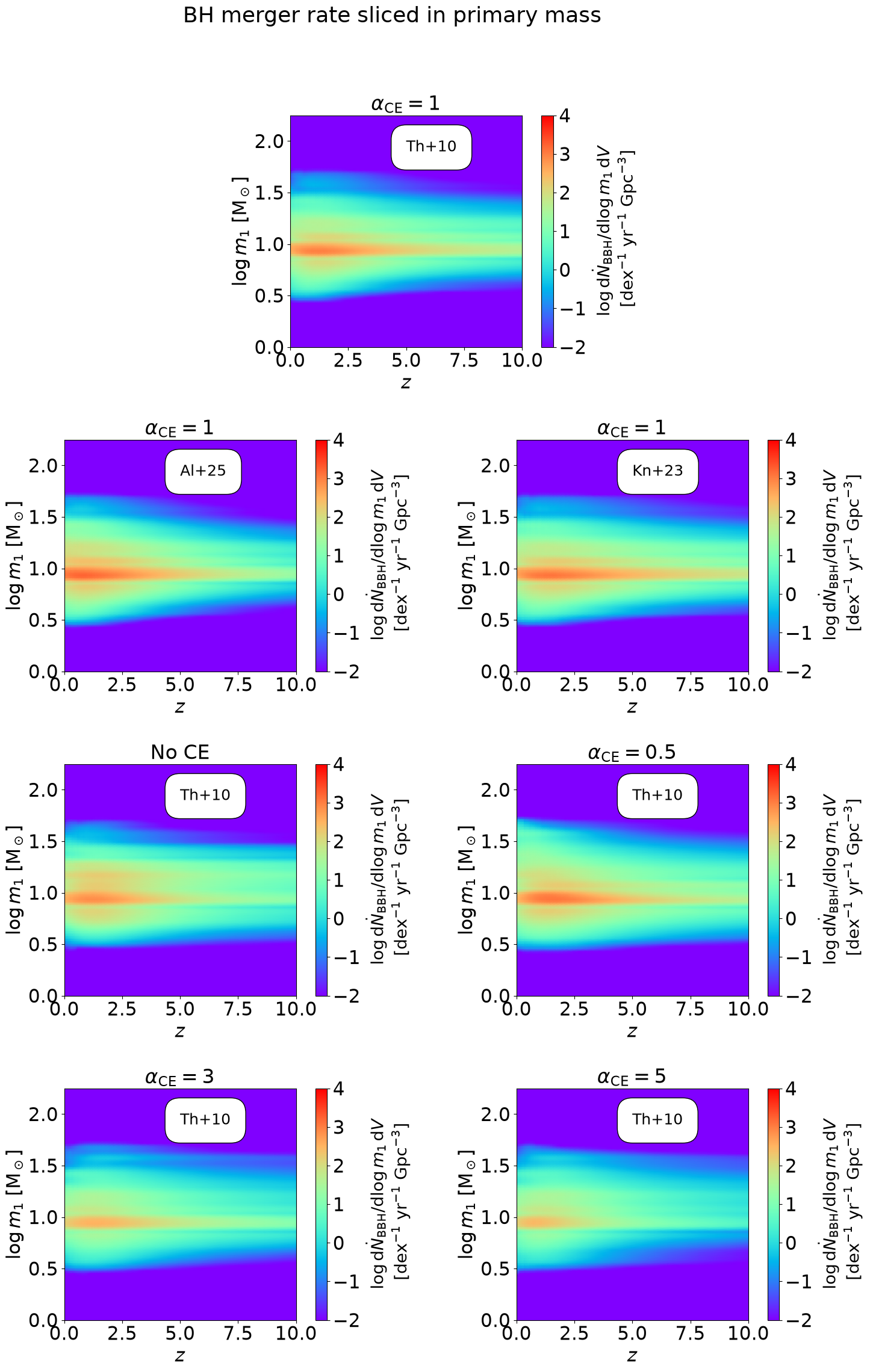}
\caption{Cosmic merger rate density of BBHs sliced in primary mass and redshift. Panels as in Figure \ref{fig|BBHMR_2D_Z}.}\label{fig|BBHMR_2D_m1}
\end{figure}

\begin{figure}[t!]
\centering\includegraphics[width=0.75\textwidth]{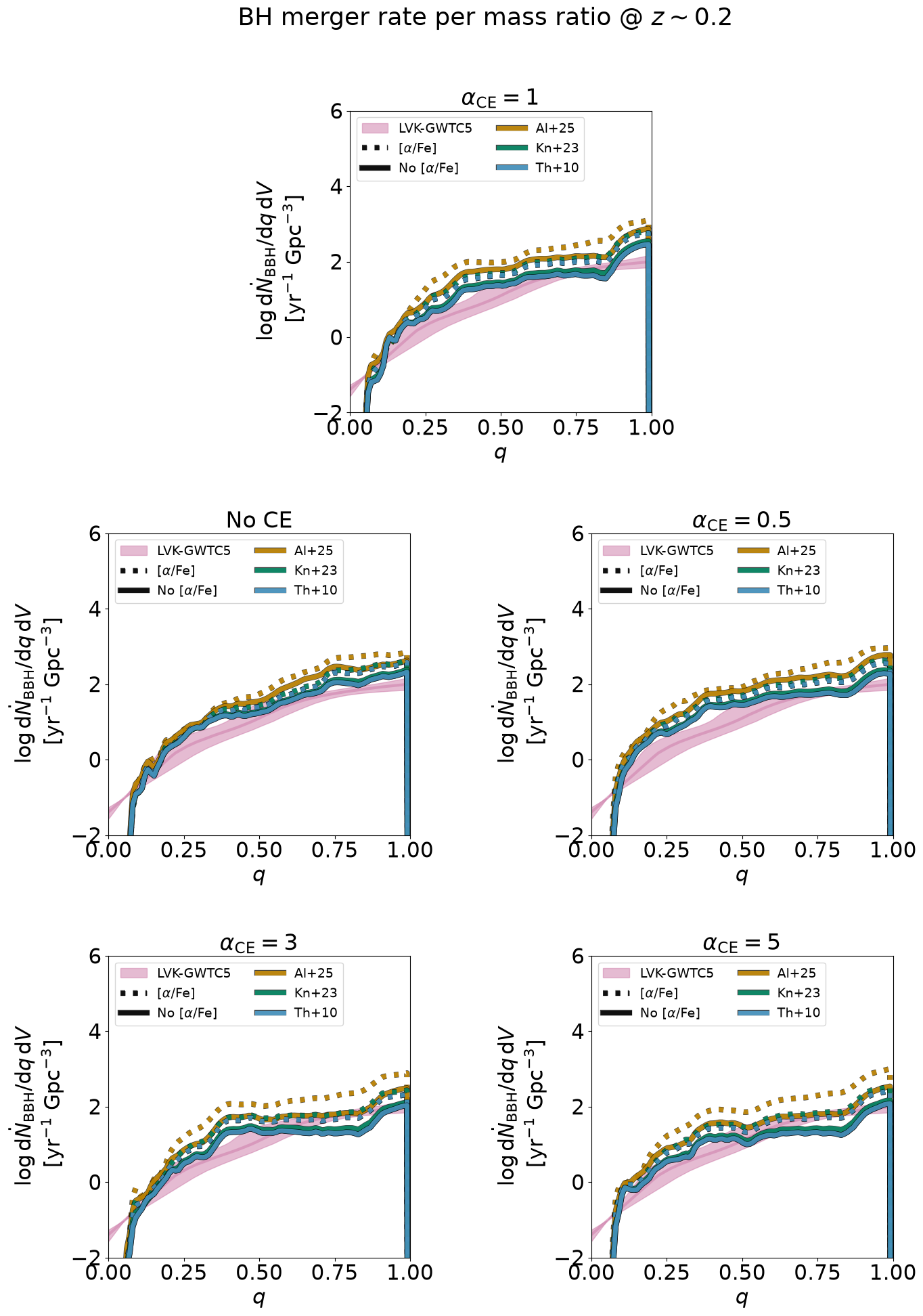}
\caption{Mass ratio distribution of BBH mergers at $z\approx 0.2$. Panels, colors and linestyles as in  Figure \ref{fig|BBHMR}.}\label{fig|BBHMR_q}
\end{figure}

\begin{figure}[t!]
\centering\includegraphics[width=0.75\textwidth]{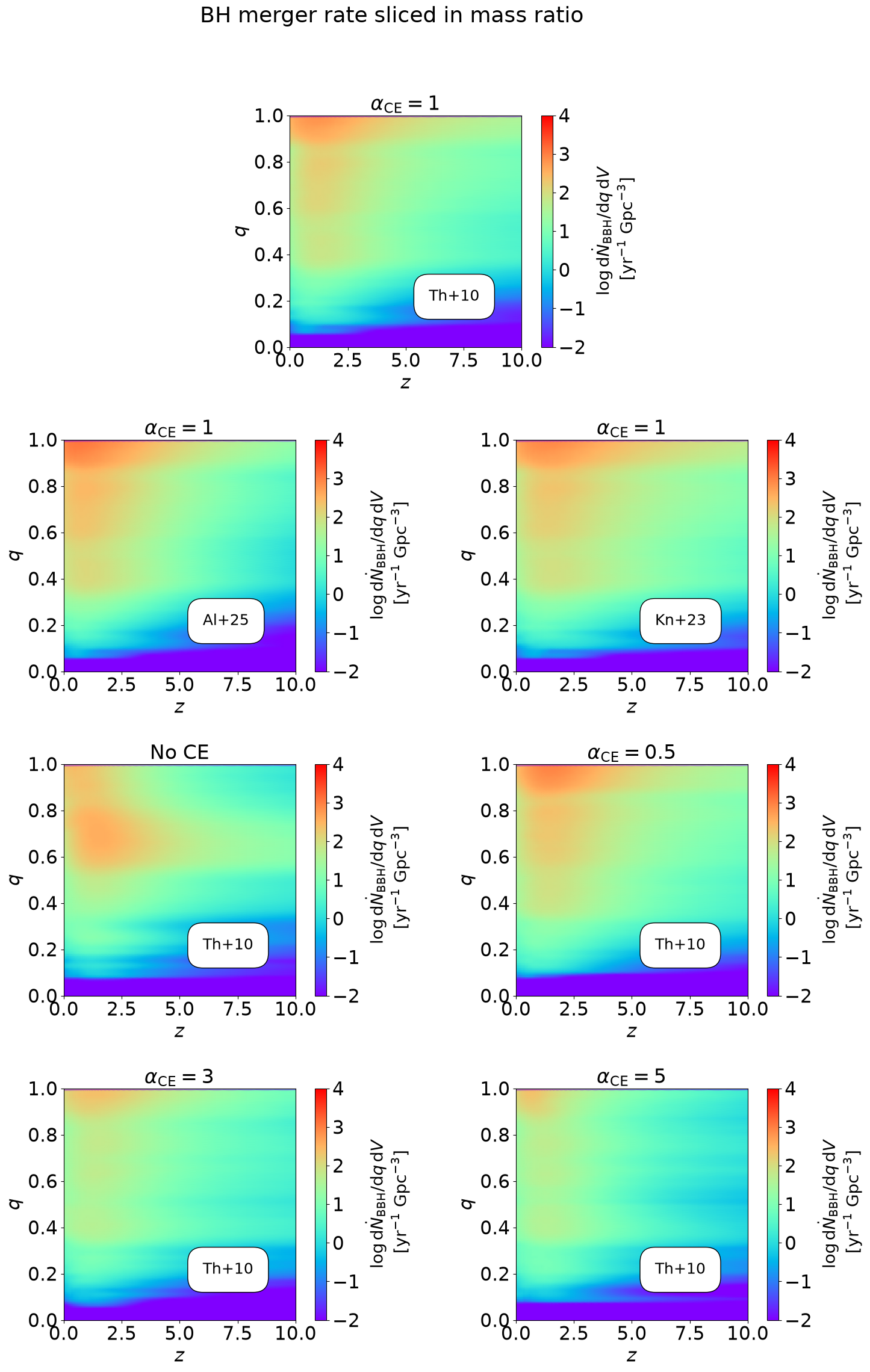}
\caption{Cosmic merger rate density of BBHs sliced in mass ratio and redshift. Panels as in Figure \ref{fig|BBHMR_2D_Z}.}\label{fig|BBHMR_2D_q}
\end{figure}

\subsection{BBH primary mass and mass ratio}
\label{sec|results_masses}

Figure~\ref{fig|BBHMR_m1} shows the local BBH merger-rate density, evaluated at $z\approx0.2$, as a function of primary mass. The analogous distributions in secondary mass and chirp mass, which display qualitatively similar trends, are reported for completeness in Appendix~\ref{app|aux}. The isolated-binary calculations predict a primary mass distribution with a pronounced concentration at $m_{\bullet,1}\sim 7-10\,M_\odot$, followed by a broader shoulder extending to a few tens of solar masses. The small-scale features visible in the curves trace the remnant-mass prescription, the metallicity dependence of stellar winds, and the treatment of pulsational pair instability and pair instability in \texttt{SEVN}.

For the comparison with GW observations, we use the GWTC-5.0 \textit{Default-BBH} population reconstruction \cite{LIGO26rec} and the \texttt{Vamana} mixture-model reconstruction \cite{Tiwari2026}, evaluated using the same source-frame mass variables and redshift convention adopted for the theoretical predictions. Comparison with such population reconstructions reinforces the conclusion derived from the integrated rates. In fact, models with No CE, $\alpha_{\rm CE}=0.5$, and $\alpha_{\rm CE}=1$ tend to overpredict the abundance of systems around the main low-mass peak. Values $\alpha_{\rm CE}=3-5$ reduce this excess and provide a closer normalization over the mass range in which the isolated channel contributes most strongly. The stellar-archaeology prescriptions mainly change the amplitude, while the detailed positions of the peaks are set by the stellar and binary evolution model. The iron-abundance correction increases the rate and slightly broadens the contribution toward larger masses.

The adopted isolated-binary model qualitatively reproduces the broad location of the low- and intermediate-mass support in the LVK population reconstruction, while providing substantially less support at the highest masses. In fact, the predicted isolated-binary distributions decline rapidly above $\sim 40-50\,M_\odot$ whereas the observational reconstructions retain a significant high-mass tail. This discrepancy is a shape issue in addition to the normalization tension, and motivates the inclusion of dynamically assembled binaries, as discussed below. 

At the low-mass end, the distributions decline sharply below the main peak, although somewhat less steeply than in the LVK population reconstruction. This feature primarily reflects the \emph{death matrix} prescription for compact-remnant masses adopted in \texttt{SEVN}, rather than a limitation of the \texttt{StAGE} framework itself. LVK detections have indeed revealed a couple of putative black holes within the lower mass gap \cite{Abbott2020_lmg,Abac24_lmg}, a region that is still sparsely sampled and may therefore be imperfectly represented by current population-level reconstructions. Future LVK observing runs and detector upgrades should substantially increase the number of sources detected in this mass range, allowing the low-mass tail of the distribution to be reconstructed more robustly and providing a sensitive test of the remnant-mass prescriptions implemented in population-synthesis models.

The corresponding redshift-dependent distribution in Figure~\ref{fig|BBHMR_2D_m1} shows that the characteristic mass scale evolves only mildly. The dominant ridges remain close to masses $\lesssim 7-10\,M_\odot$ over a broad redshift range. The main evolution is in the normalization, which follows the cosmic merger history, together with a gradual narrowing of the high-mass support toward the highest redshifts. This weak mass evolution arises because the BBHs merging at a given epoch sample a mixture of birth metallicities and delay times, which partly washes out the stronger metallicity dependence visible in the underlying \texttt{SEVN} kernels.

Figures~\ref{fig|BBHMR_q} and~\ref{fig|BBHMR_2D_q} show the corresponding mass-ratio distributions. The isolated channel strongly favors comparable component masses: the rate rises toward $q_\bullet\sim 1$, systems with $q_\bullet\gtrsim 0.5$ dominate at every redshift, and very asymmetric binaries with $q_\bullet\lesssim 0.2$ are strongly suppressed. The overall shape is broadly compatible with the LVK reconstruction, although the normalization is again high for the lower values of $\alpha_{\rm CE}$. The preference for nearly equal masses persists with redshift, while the detailed curvature at intermediate $q_\bullet$ retains a moderate dependence on the CE prescription.

\subsection{Impact of dynamical evolution in dense environments}
\label{sec|results_dyn}

\begin{figure}[t!]
\centering\includegraphics[width=0.825\textwidth]{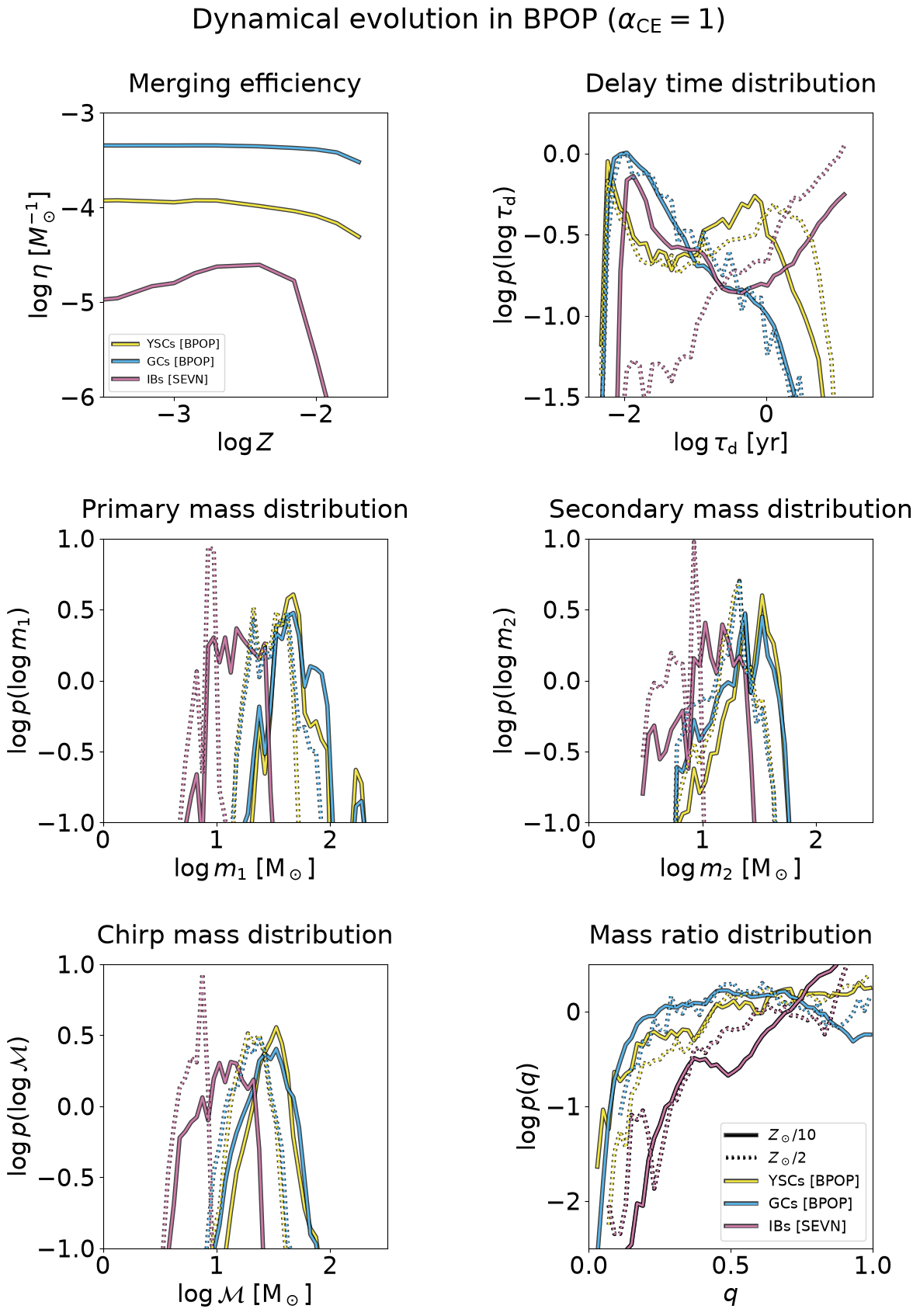}
\caption{Properties of the pre-computed dynamical BBH catalogs generated with the \texttt{BPOP} framework: merger efficiency as a function of metallicity (top left), delay time distribution (top right), primary mass distribution (middle left), secondary mass distribution (middle right), chirp mass distribution (bottom left), and mass ratio distribution (bottom right). The reference common envelope parameter $\alpha_{\rm CE} = 1$ has been adopted. 
Colored lines refer to isolated binaries (magenta; same as Figure \ref{fig|SEVN}), to binaries in young star clusters (yellow) and in globular clusters (cyan). In all panels but the top left, solid lines refer to metallicity $Z_\odot/10$ and dotted lines to $Z_\odot/2$.}\label{fig|BPOP}
\end{figure}

Figure~\ref{fig|BPOP} compares the merger kernels for isolated binaries with those derived from the \texttt{BPOP} models of YSCs and GCs. The merger efficiency per unit stellar mass formed in the relevant environment is substantially larger in dense clusters: the GC kernel is the most efficient, the YSC kernel is intermediate, and the isolated-binary kernel is the least efficient over most of the metallicity range. The cosmic importance of these channels, however, also depends on the much smaller fraction of stellar mass assigned to bound clusters in our normalization. Another relevant feature is that, while the efficiency for the isolated channel tends to steeply decline for metallicities $Z\gtrsim Z_\odot/2$, in dense clusters it stays quite constant out to around solar values.

Such a weak metallicity dependence of the GC merger efficiency in the adopted \texttt{BPOP} catalogs is not unique to these calculations, although it should not be regarded as a universal prediction of the dynamical channel. Several dynamical-cluster studies have found that the BBH merger efficiency depends only weakly on metallicity over the subsolar regime relevant for most GCs, since dynamical pairing and hardening can partly compensate for metallicity-dependent stellar-evolution effects \cite{Rodriguez2016,Chatterjee2017,Hong2018,Mapelli2022}. A similarly weaker metallicity dependence than in isolated binaries has been found for YSCs by \cite{DiCarlo2020}. On the other hand, other works (e.g., \cite{Rodriguez2023}) have found that higher-metallicity clusters retain fewer BHs because stronger stellar winds produce lower-mass remnants that generally receive larger natal kicks. The initial BH-retention fraction, however, does not map directly onto the number of BBH mergers per unit cluster mass, since subsequent dynamical pairing, hardening, ejection, and cluster evolution also enter the merger efficiency. The quantitative metallicity dependence therefore remains sensitive to the stellar-remnant and natal-kick prescriptions as well as to the initial and evolving cluster properties. We consequently regard the mild metallicity dependence of the \texttt{BPOP} kernels adopted here as a model-dependent, but not unprecedented, feature of the dynamical channel.

The intrinsic distributions from \texttt{BPOP} clearly show the imprint of dynamical processing. YSCs and GCs shift the mass distributions toward larger values and extend them well beyond the upper support of the isolated channel. This is a consequence of mass segregation, exchange interactions, stellar collisions, and the retention and recycling of merger remnants. The dynamical mass-ratio distributions are also broader and less strongly concentrated toward $q_\bullet\sim 1$, because exchange interactions can pair black holes that did not share a common stellar binary origin. The delay time distributions differ from the isolated case and reflect the internal cluster-evolution timescale and the subsequent hardening or ejection of BBHs.

\begin{figure}[t!]
\centering\includegraphics[width=0.75\textwidth]{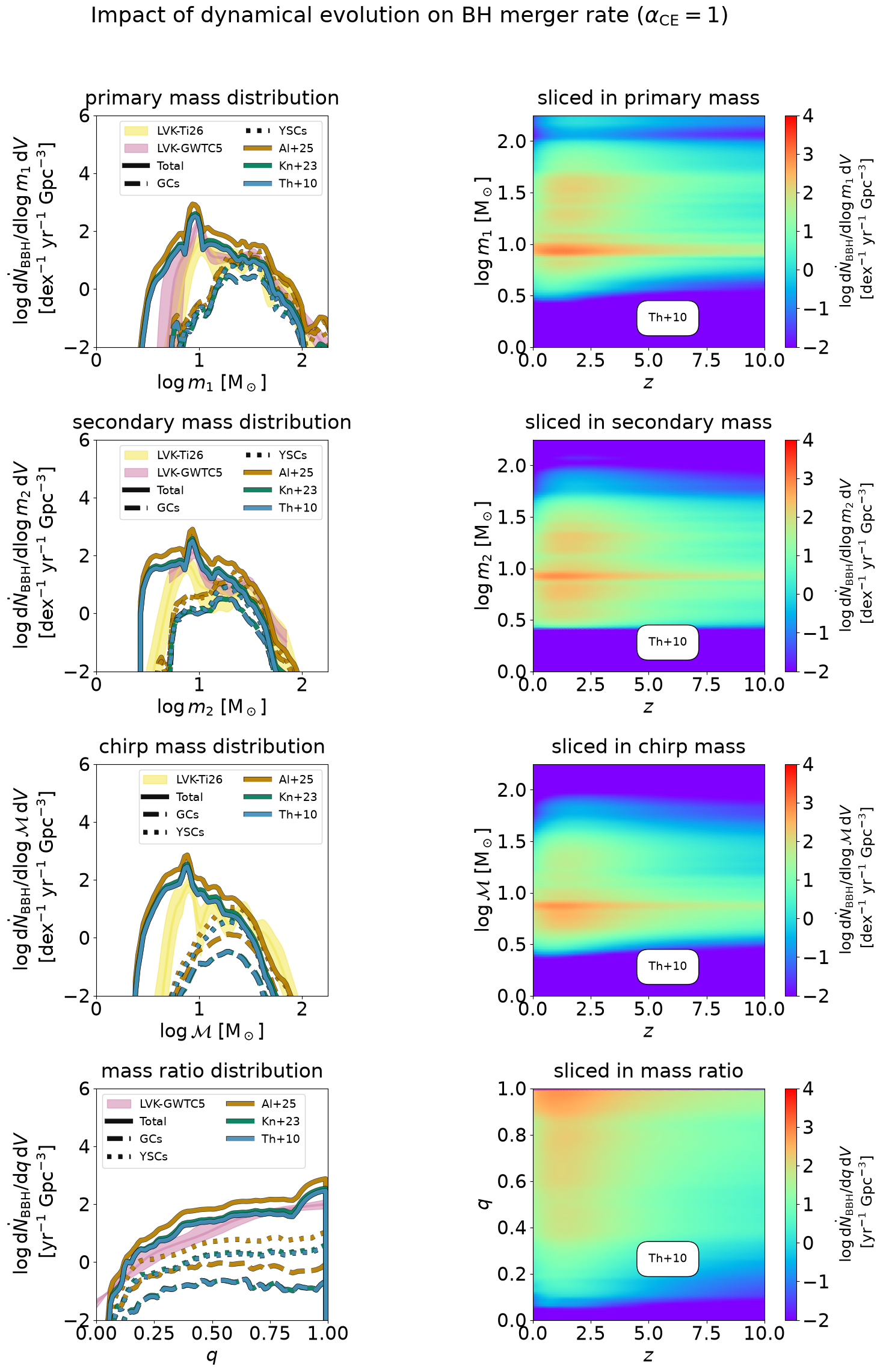}
\caption{Impact of dynamical evolution in dense environments on the BBH merger rate density as a function of primary mass (first row),
secondary mass (second row), chirp mass (third row), and mass ratio
(fourth row). In the left column, colored solid lines show the total
isolated+YSC+GC rates at $z\approx 0.2$ for the three stellar-archaeology
prescriptions. The corresponding dashed and dotted lines show the GC
and YSC contributions, respectively. The right column shows the total
merger rate density, color coded as a function of redshift and the
corresponding binary parameter, for the \citetalias{Thomas2010}
prescription. The common envelope efficiency is $\alpha_{\rm CE}=1$ throughout.}\label{fig|BBHMR_dyn}
\end{figure}

The impact after convolution with the \texttt{StAGE} histories is displayed in Figure~\ref{fig|BBHMR_dyn}. With the adopted YSC and GC normalizations, the total rate remains dominated by isolated binaries around the main low-mass peak. The dynamical channels become progressively more important toward larger component and chirp masses. In particular, they fill the region above $m_{\bullet,1}\sim40-50\,M_\odot$ and extend the total distribution toward $m_{\bullet,1}\gtrsim100\,M_\odot$, providing additional support at high masses and thereby improving the qualitative agreement with the LVK population reconstruction. A similar extension is visible in the secondary- and chirp-mass distributions. The redshift-resolved panels show that this high-mass component is not confined to the local Universe but persists over a broad range of merger epochs.

The dynamical contribution also adds support at intermediate and low mass ratios. Nevertheless, for the fiducial cluster normalizations adopted here, the total mass-ratio distribution remains dominated by the isolated component and continues to rise toward equal masses. Hence, the clearest signature of the cluster channels in the present calculation is the extension of the mass distributions rather than a dramatic reshaping of the total $q_\bullet$ distribution. Systems with more asymmetric masses may be more naturally produced in some dynamical scenarios, although the mass ratio alone does not provide a unique formation-channel diagnostic.

For the fiducial YSC and GC normalizations adopted here, the dynamical channels populate the high-mass end of the distribution while also increasing the total merger-rate density. This exacerbates the normalization discrepancy with the LVK-inferred range. These results illustrate that a mixed-channel model could alleviate the mismatch in the mass distribution, although a quantitative calibration of the relative isolated and dynamical contributions is beyond the scope of the present analysis. A successful mixed-channel model will likely require a reduced effective efficiency for the isolated population together with a calibrated dynamical contribution that supplies the high-mass tail, consistently with the broader picture explored in \cite{Zevin2021,Arcasedda2026} and with the recent, partly model-independent analysis by \cite{Bosi2026}.

\begin{figure}[t!]
\centering\includegraphics[width=0.8\textwidth]{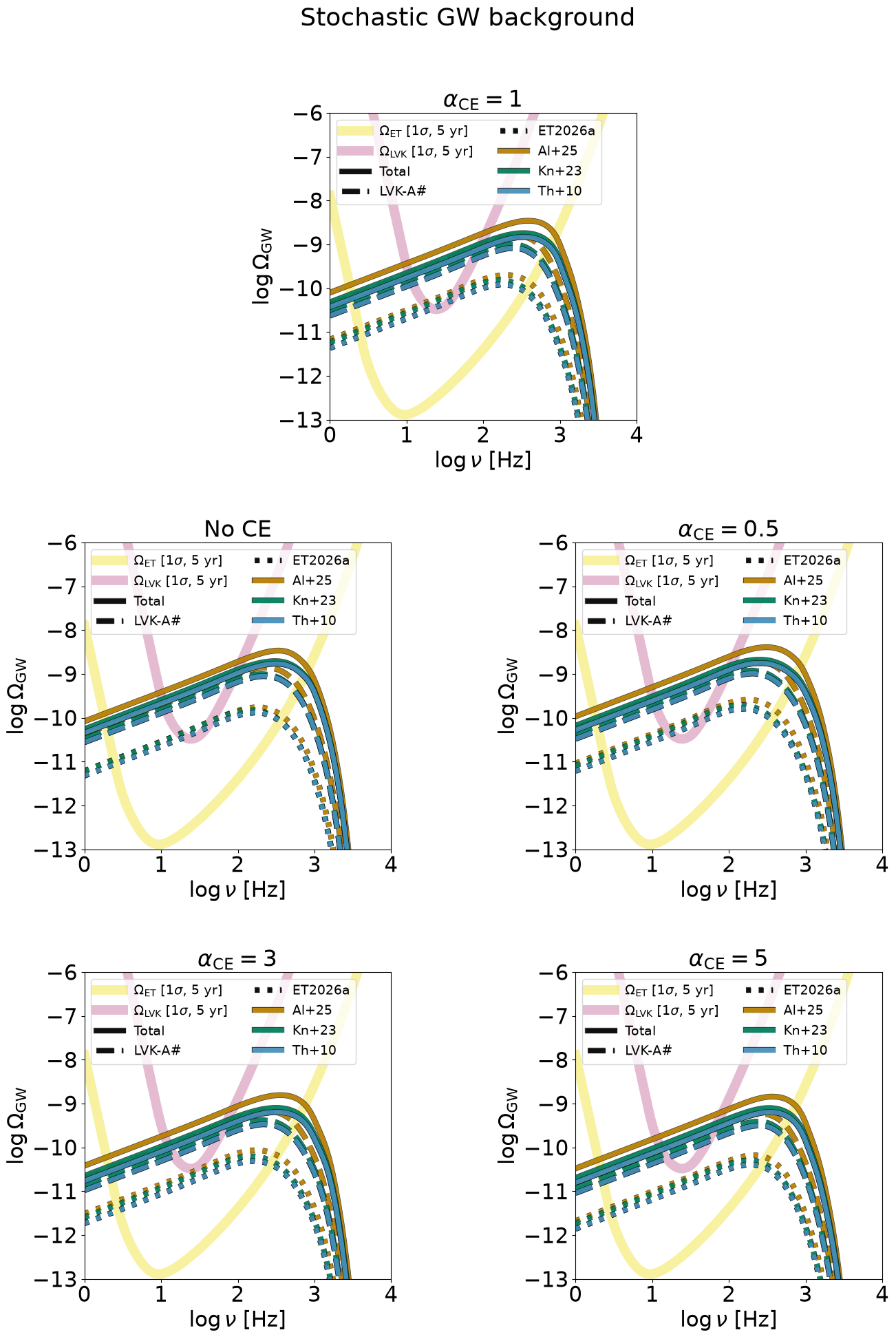}
\caption{Stochastic GW background from BBH mergers in the isolated-binary channel. Different panels refer to the reference model
$\alpha_{\rm CE}=1$ (top), the No CE model (middle left), and
$\alpha_{\rm CE}=0.5$, $3$, and $5$ (middle right, bottom left, and
bottom right, respectively). Colors identify the
\citetalias{Thomas2010} (blue), \citetalias{Alvarez2025} (orange), and
\citetalias{Knowles2023} (green) stellar-archaeology prescriptions. Solid lines
show the total background. Dashed and dotted lines show the unresolved
backgrounds obtained after removing events with ${\rm S/N}>12$ for the
LVK A\# and ET2026a configurations, respectively. Thick magenta and yellow
curves show the five-year, $1\sigma$ power-law integrated sensitivities
for LVK A\# and ET2026a.}\label{fig|SGWB}
\end{figure}

\subsection{Stochastic GW background}
\label{sec|results_sgwb}

Figure~\ref{fig|SGWB} shows the SGWB generated by
the isolated BBH population (we checked that the inclusion of the dynamical channel has a minor impact). The total spectra exhibit the expected inspiral-dominated rise at low frequency and a turnover at a few hundred Hz as the merger and ringdown portions of the source-frame spectra enter the observed band. In the frequency range relevant for ground-based detectors, the predicted amplitudes are typically of order $\Omega_{\rm GW}\sim10^{-10}-10^{-9}$, with differences of a factor of a few among the stellar archaeology models, reflecting the different amounts and timing of star formation. Increasing $\alpha_{\rm CE}$ reduces the amplitude, consistently with the lower merger rate normalization.

For each detector, the unresolved spectrum is obtained after removing events with ${\rm S/N}>12$. The unresolved background for the LVK A\# configuration remains close to the total one, because many high-redshift mergers are still individually undetectable. By contrast, ET2026a resolves a much larger fraction of the BBH population, lowering the residual background by roughly an order of magnitude over much of the band. Even so, the much greater stochastic sensitivity of ET compensates for this reduction.

At the adopted benchmark of observing time around $T=5$ years and at $1\sigma$ PI sensitivity, the LVK A\# unresolved spectra generally reach or cross the LVK sensitivity curve, although the margin is smallest for the models with $\alpha_{\rm CE}=3-5$ that have a lower normalization. The ET residual backgrounds lie comfortably above the ET2026a sensitivity curve for all the CE prescriptions considered. Therefore, the upgraded LVK network can provide a meaningful test of the higher-amplitude models, while ET should be sensitive to both the integrated BBH background and the residual left after the subtraction of individually resolved sources.

The SGWB provides information complementary to the local merger rate and mass distributions. It integrates the unresolved population over cosmic time and is consequently sensitive to the extended high-redshift tail, where the stellar-archaeology prescriptions differ more strongly. Models that predict a large resolved merger rate density also tend to generate a larger background. Jointly fitting the resolved population and the SGWB can therefore help separate uncertainties associated with galaxy star formation and CE treatment or other binary evolution prescriptions.

\section{Discussion and conclusions}
\label{sec|summary}

In this work we have applied the stellar archaeology driven framework \texttt{StAGE} to the formation and mergers of BBHs associated with present-day quiescent galaxies and their star-forming progenitors. The star formation and chemical-enrichment histories inferred for the progenitors of present-day QGs were convolved with merger kernels from \texttt{SEVN} for isolated binaries and, in an exploratory extension, with dynamical kernels from \texttt{BPOP} for dense stellar environments. Unlike cosmic-average approaches, the star formation histories entering \texttt{StAGE} are reconstructed backward from the local QG mass function and stellar-archaeology constraints. The distinctive novel contribution of the present framework is twofold: it preserves the connection between the environment in which BBHs form and the properties of the galaxy in which they eventually merge, and it exploits the measured $\alpha$-enhancement of QGs to investigate the iron abundance relevant for massive-star evolution rather than relying only on oxygen-based total metallicity. Moreover, some of our other results, in particular the high normalization obtained for several isolated-binary prescriptions and the extension of the high-mass support of the primary mass distribution by dynamical evolution, provide an independent test of conclusions already emerging from recent literature studies.

The separation between binary birth and merger environments is a particularly useful feature of \texttt{StAGE}. The SFR and metallicity relevant for stellar evolution must be evaluated at binary formation, whereas the stellar mass, burstiness, and other properties relevant for host identification must be evaluated at coalescence. Most BBHs are born at moderately subsolar metallicity, but their hosts evolve from still-growing systems at high redshift to massive descendants at late cosmic times. Around $z\sim1-2$, where the merger-rate density is largest, the dominant hosts lie on or above the star-forming main sequence, whereas toward the local Universe long delay times generate an increasing contribution from galaxies that have entered the quenching or quiescent regime. A BBH merging in a passive galaxy therefore need not have formed in a passive environment, but may instead represent the delayed outcome of the intense star-forming episode that assembled the QG \cite{Artale2019,Santoliquido2022,Srinivasan2023}. The detailed partition among starbursting, main-sequence, and quenching hosts depends appreciably on the adopted stellar-archaeology prescription, while being comparatively insensitive to $\alpha_{\rm CE}$, illustrating the complementary roles of galaxy and binary evolution in shaping the observable BBH population.

This evolutionary information may also be valuable for GW cosmology. Dark-siren analyses commonly weight candidate hosts using observables such as luminosity or stellar mass \cite{Schutz1986,DelPozzo2012,Fishbach2019,Gray2020,Mastrogiovanni2024}. \texttt{StAGE} provides instead redshift-dependent priors in stellar mass and star-formation state that, for the BBH population associated with QG progenitors, are tied to their reconstructed formation histories and delay-time distribution. Such information could be incorporated into statistical host-association and dark-siren analyses; assessing the actual gain will require convolving the predicted distributions with the selection function and completeness of the adopted galaxy catalog.

The comparison between total metallicity and iron abundance highlights the second distinctive aspect of the present analysis. The birth metallicities of the merging systems are concentrated around $Z\sim Z_\odot/10-Z_\odot/3$, where line-driven winds remain sensitive to the detailed abundance pattern. Since QG progenitors are $\alpha$-enhanced, their iron abundance can be appreciably lower than the one inferred from their total metal content, which is commonly traced by oxygen. Our approximate treatment based on $Z_{\rm Fe}$ indicates that abundance-ratio effects can appreciably modify the BBH merger efficiency when binary-evolution models are applied to rapidly formed, $\alpha$-enhanced stellar populations. In the present calculations the lower effective iron abundance increases the predicted merger rate, although a fully self-consistent assessment will require dedicated stellar-evolution calculations with the appropriate abundance mixtures \cite{Chruslinska2024b,Chruslinska2025,Romagnolo2026,Boco2026a}.

Using this independent galaxy-evolution construction, we also recover the tendency of several isolated-binary models to predict BBH merger rates above the LVK-inferred range \cite{Broekgaarden2022,Santoliquido2022,Srinivasan2023,Romagnolo2023,Boesky2024,Sgalletta2025,Boco2026a,Broekgaarden2026, Boco2026b}. Although the progenitors of present-day QGs account for only a subset of cosmic star formation, the standard CE prescriptions already produce a local merger-rate density exceeding the observational estimates. This conclusion is robust against the adopted stellar-archaeology prescription and is strengthened when the $\alpha$-enhanced iron abundance is used in place of the oxygen-based metallicity. Within the specific \texttt{SEVN} grid explored here, the predictions approach the observed normalization only for the largest effective values $\alpha_{\rm CE}\gtrsim3$, or for an equivalent reduction in the merger-kernel normalization. This should not be interpreted as evidence that such extreme CE efficiencies are physically favored, nor as implying a universal failure of isolated binary evolution: predictions across the literature span a broad range as stellar, binary, and cosmic-evolution assumptions are varied \cite{Broekgaarden2022,Santoliquido2022,Sgalletta2025,Boco2026a,Broekgaarden2026,Boco2026b}. Rather, \texttt{StAGE} provides an independent, observationally anchored route to the same conclusion that the ingredients controlling the effective merger efficiency of the isolated channel remain a major source of uncertainty and may need to be revisited.

A similar distinction between new and confirmatory results applies to the binary-mass distributions. Consistently with previous population-synthesis studies, isolated evolution produces substantial support over the low- and intermediate-mass ranges and favors relatively comparable component masses, but its predicted distributions decline rapidly toward the highest primary and chirp masses. Dynamical processing in YSCs and GCs naturally extends the distributions beyond the isolated-binary cutoff through exchange interactions, stellar collisions, and hierarchical mergers \cite{Mapelli2020,Torniamenti2024,Arcasedda2026}. Within the \texttt{StAGE} framework, however, adding this dynamical contribution also increases the total merger-rate density and therefore somewhat exacerbates the normalization discrepancy. Our results consequently motivate a mixed picture in which the effective isolated-binary efficiency is reduced while dynamical evolution provides a smaller but important contribution to the most massive systems. Quantifying this balance will require a joint inference of the isolated and dynamical channels.

The SGWB provides an independent cumulative test of the same physical ingredients. For much of the parameter space considered here, the predicted unresolved spectra approach the five-year, $1\sigma$ sensitivity of the planned LVK A\# configuration and lie comfortably above the corresponding ET2026a sensitivity over a broad frequency range. ET will individually resolve a much larger fraction of the BBH population, but the residual unresolved background remains informative because of the detector's substantially improved sensitivity. Jointly analyzing the redshift-dependent merger rate, intrinsic mass distributions, host-galaxy properties, and SGWB will be especially powerful: resolved catalogs constrain the nearby and individually detectable population, whereas the background retains sensitivity to the numerous unresolved systems at higher redshift, where uncertainties in galaxy star formation and binary delay-time distributions become increasingly important \cite{Callister2020,Turbang2024,Dvorkin2016b,Cousin2026}.

A natural future development is to extend the same stellar-archaeology framework from stellar-mass to massive black holes. Coupling the reconstructed galaxy star formation histories to AGN Eddington ratio distribution and auxiliary constraints on black-hole accretion would provide growth histories for the central massive black holes of QGs and their progenitors, opening the possibility of predicting massive-BH binary populations, their nanohertz GW background, and individually detectable \textit{LISA} sources. The same environmental histories could also be exploited to investigate additional compact-object channels, including mergers in AGN disks and tidal-disruption events.

In summary, \texttt{StAGE} provides an observationally anchored bridge between the fossil record of present-day galaxies and the cosmic population of merging compact binaries. Its main added value is the ability to jointly connect binary birth conditions, abundance patterns, merger-time host properties, and GW observables within the same data-driven galaxy evolution framework. Hopefully \texttt{StAGE} will help in paving the way toward the multiscale perspective that will be essential for advancing GW astrophysics and cosmology in the near future.

\begin{tcolorbox}[colback=gray!8,colframe=black,title={Take-home messages}]

\begin{itemize}

\item \texttt{StAGE} separately tracks BBH birth conditions and merger-time host properties. Most BBHs form at moderately subsolar metallicities, while their hosts evolve in stellar mass and star-formation state. This provides redshift-dependent priors for host galaxy association and dark-siren cosmology.

\item \texttt{StAGE} shows that the abundance pattern can appreciably affect BBH merger rates. Specifically, using stellar-archaeology constraints on $\alpha$-enhancement, the effective iron abundance relevant for massive-star winds is lower than the total metallicity, increasing BBH merger rates.

\item The star-forming progenitors of present-day QGs can contribute substantially to the cosmic BBH population, with a merger-rate density peaking at $z\approx1-2$ and extending to higher redshift.

\item Using the independent \texttt{StAGE} approach, we confirm that several binary-evolution prescriptions overpredict the local LVK-inferred rate. Agreement improves for extreme values $\alpha_{\rm CE}\gtrsim 3$ or, more generally, for a reduced isolated-binary merger efficiency.

\item Exploiting \texttt{StAGE} we also confirm that dynamical formation in stellar clusters extends the BBH primary mass distribution into the high-mass regime poorly populated by isolated binaries.

\item The predicted SGWB lies within the projected reach of LVK A\# and above the five-year, $1\sigma$ ET2026a sensitivity, providing a complementary probe of unresolved high-redshift mergers and galaxy vs. stellar binary evolution degeneracies.

\end{itemize}

\end{tcolorbox}

\acknowledgments
We thank the anonymous referee for helpful and constructive comments and suggestions. II and AL have been supported by the Istituto Nazionale di Fisica Nucleare (INFN) via the specific national initiative QGSKY. MS and GA acknowledge financial support from the Istituto Nazionale di Fisica Nucleare (INFN), Sezione di Trieste, through the ET-Italia and TEONGRAV initiatives. MS acknowledges financial support from the Istituto Nazionale di Fisica Nucleare (INFN), Sezione di Trieste, through the VIRGO initiative. MS acknowledges support from the INAF-Large Grant 2024: ``Envisioning Tomorrow: prospects and challenges for multimessenger astronomy in the era of Rubin and Einstein Telescope''. 
CS acknowledges financial support from the Alexander von Humboldt Foundation for the Humboldt Research Fellowship. LB acknowledges support by the Deutsche Forschungsgemeinschaft (DFG, German Research Foundation) in the form of a Walter Benjamin position – Projektnummer 555003977.
LB and CS acknowledge support from the German Excellence Strategy via the Heidelberg Cluster of Excellence (EXC2181 - 390900948) STRUCTURES.
MB acknowledges that this article was produced while attending the PhD program in PhD in Space Science and Technology at the University of Trento, Cycle XXXIX, with the support of a scholarship financed by the Ministerial Decree no. 118 of 2nd March 2023, based on the NRRP - funded by the European Union - NextGenerationEU - Mission 4 "Education and Research", Component 1 "Enhancement of the offer of educational services: from nurseries to universities” - Investment 4.1 “Extension of the number of research doctorates and innovative doctorates for public administration and cultural heritage” - CUP E66E23000110001 and support by the Italian grant Project SPACE-IT-UP by the Italian Space Agency and Ministry of University and Research, Contract Number 2024-5-E.0. 

\bigskip

This research has made use of data or software obtained from the Gravitational Wave Open Science Center (gwosc.org), a service of the LIGO Scientific Collaboration, the Virgo Collaboration, and KAGRA.

\appendix

\section{BBH merger rates for the overall galaxy population}\label{app|cosmorate}

We recall here the standard data-driven approach to compute the BBH merger rates for the overall galaxy populations \cite{Dominik2013,Belczynski2016,Neijssel2019, Chruslinska2019,Santoliquido2020,Santoliquido2022,Sgalletta2025}. The starting point is the classic analytic fit to the overall cosmic SFR density estimated from UV and IR galaxy surveys \cite{Madau2014rev}:
\begin{equation}
\rho_{\rm SFR}(z) = 0.01\,\frac{(1+z)^{2.7}}{1+[(1+z)/2.9]^{5.6}}~~M_{\odot}~{\rm yr}^{-1}~{\rm Mpc}^{-3}~,
\end{equation}
where the normalization has been rescaled from the Salpeter to the Kroupa IMF adopted here. This is then supplemented with a log-normal distribution in metallicity  
\begin{equation}
\frac{\text{d}p}{\text{d}\log Z}(Z,z) = \frac{1}{\sqrt{2\pi}\,{\Sigma}_{\log Z}}\, \exp\left[-\frac{[\log(Z/Z_\odot)-\langle\log (Z/Z_\odot)\rangle]^2}{2\,{\Sigma}_{\log Z}^2}\right]
\end{equation}
where $\langle\log (Z/Z_{\odot})\rangle \approx 0.05 - 0.14\, (1+z)$ is the redshift-dependent mean value and $\Sigma_{\log Z}\approx 0.4$ dex is the dispersion. This can be considered a fair representation of the measured oxygen-based metallicity averaged over the global galaxy populations at a given redshift  (e.g., see Figure 3 in \cite{Chruslinska2025} and references therein). If one wishes to rely on the iron-based metallicity, then the mean relation should be modified to read $\langle\log (Z_{\rm Fe}/Z_{\odot})\rangle \approx -0.24 - 0.16\, (1+z)$. 

Then the cosmic SFR density per unit metallicity just reads
\begin{equation}
\frac{\mathrm{d}\rho_{\rm SFR}}{\mathrm{d}\log Z}(Z, z) = \rho_{\rm SFR}(z)\times \frac{\text{d}p}{\text{d}\log Z}(Z,z)~,
\end{equation} 
and, once a delay time distribution has been provided, it could be exploited to compute the BBH merger rates via Equation (\ref{eq|Rmerg}) of the main text. We report in Figure \ref{fig|BBHMR} the resulting BBH rates for the overall galaxy populations as a reference. 

\section{Additional BBH mass distributions}\label{app|aux}

For completeness, this Appendix presents the BBH merger-rate distributions in secondary mass and chirp mass from \texttt{StAGE}, complementing the primary-mass and mass-ratio results of Section \ref{sec|results}. They show the same main trends: stellar archaeology mainly affects the normalization, whereas the characteristic mass scales are set by stellar and binary evolution. The isolated-binary distributions decline rapidly at the highest masses and evolve only mildly with redshift.

\begin{figure}[b!]
\centering\includegraphics[width=0.7\textwidth]{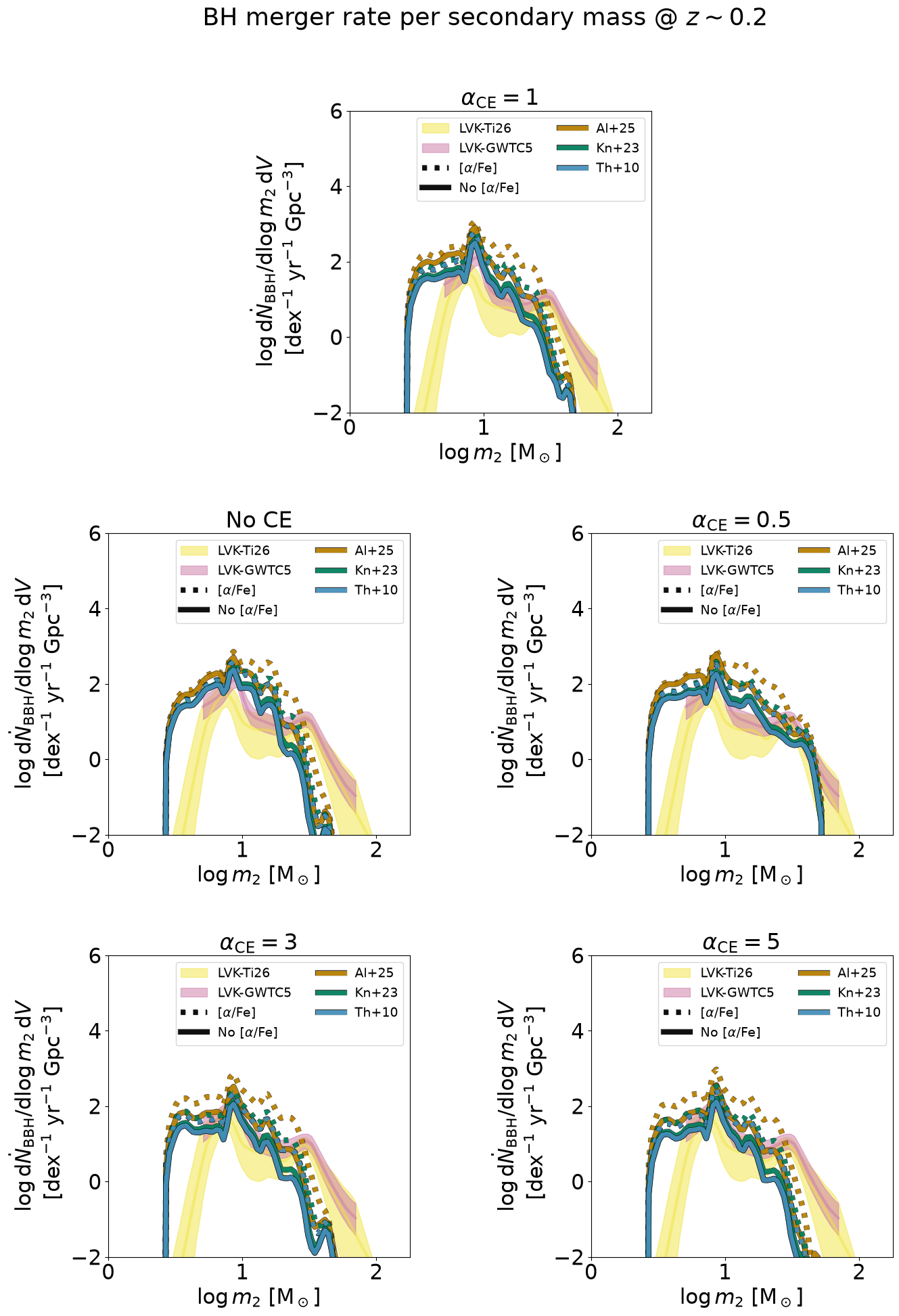}
\caption{Secondary mass distribution of BBH mergers at $z\approx 0.2$. Panels, colors and linestyles as in  Figure \ref{fig|BBHMR}.}\label{fig|BBHMR_m2}
\end{figure}

\clearpage

\begin{figure}[t!]
\centering\includegraphics[width=0.7\textwidth]{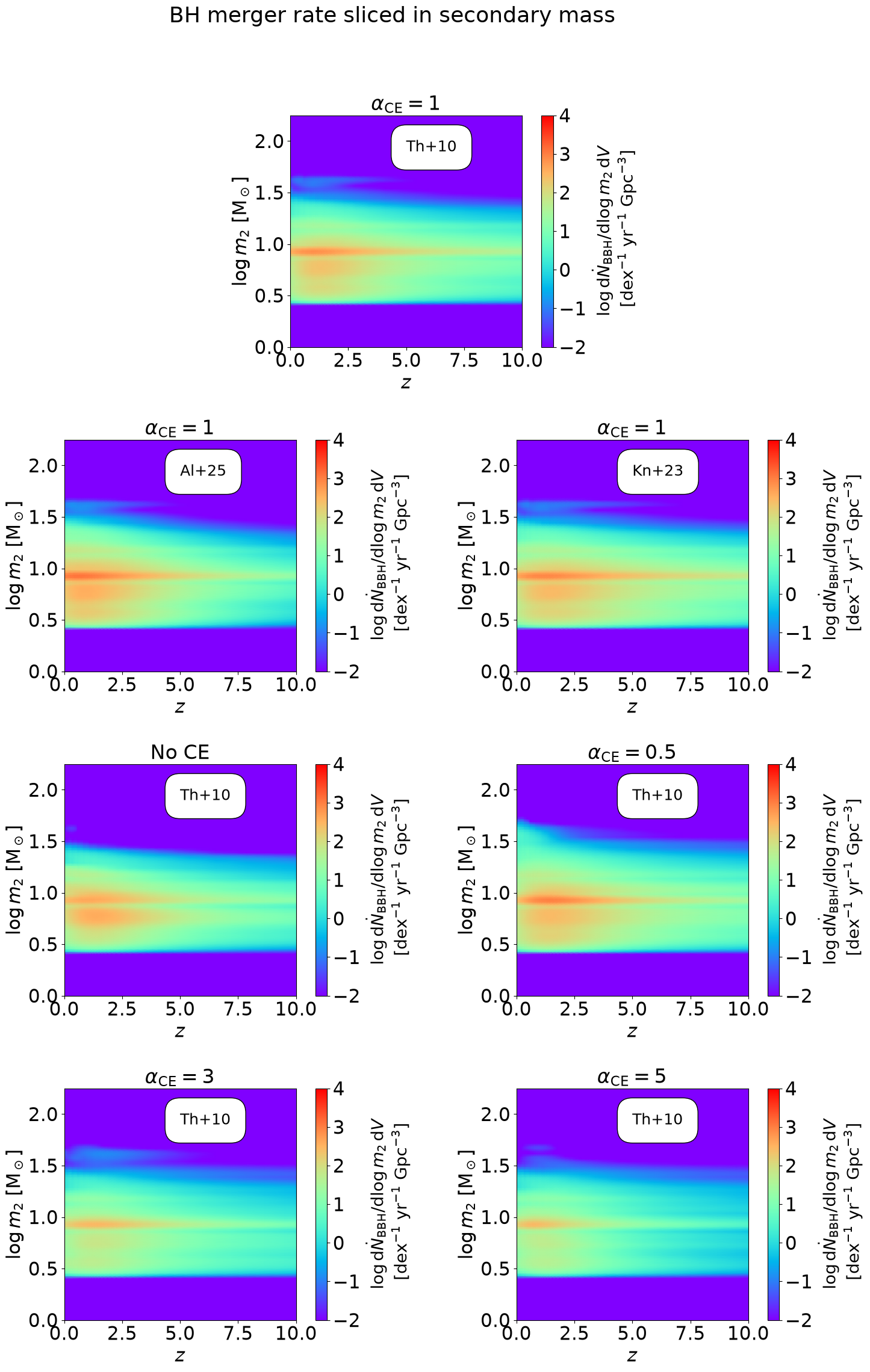}
\caption{Cosmic merger rate density of BBHs sliced in secondary mass and redshift. Panels as in Figure \ref{fig|BBHMR_2D_Z}.}\label{fig|BBHMR_2D_m2}
\end{figure}

\clearpage 

\begin{figure}[t!]
\centering\includegraphics[width=0.7\textwidth]{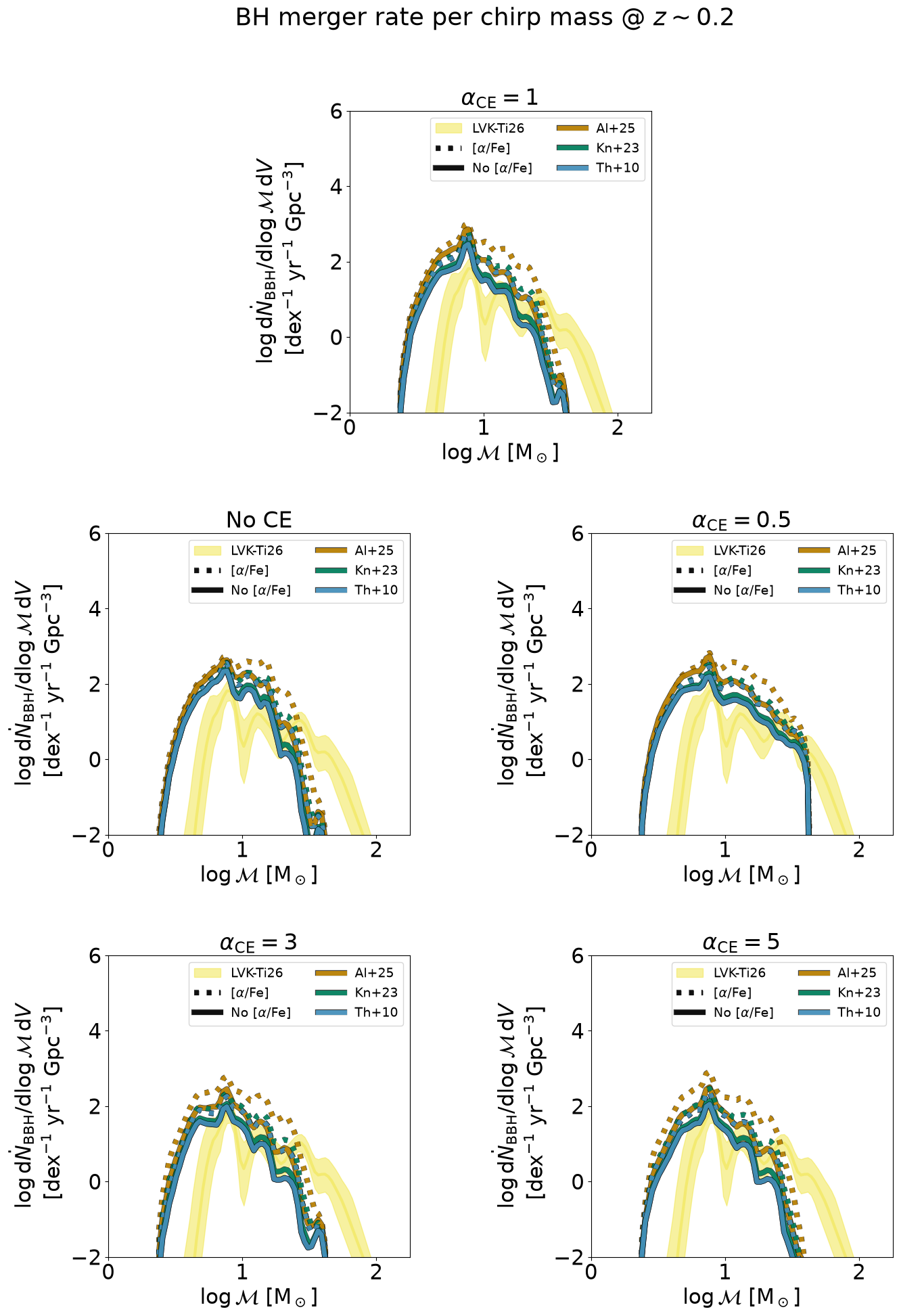}
\caption{Chirp mass distribution of BBH mergers at $z\approx 0.2$. Panels, colors and linestyles as in  Figure \ref{fig|BBHMR}.}\label{fig|BBHMR_mchirp}
\end{figure}

\clearpage

\begin{figure}[t!]
\centering\includegraphics[width=0.7\textwidth]{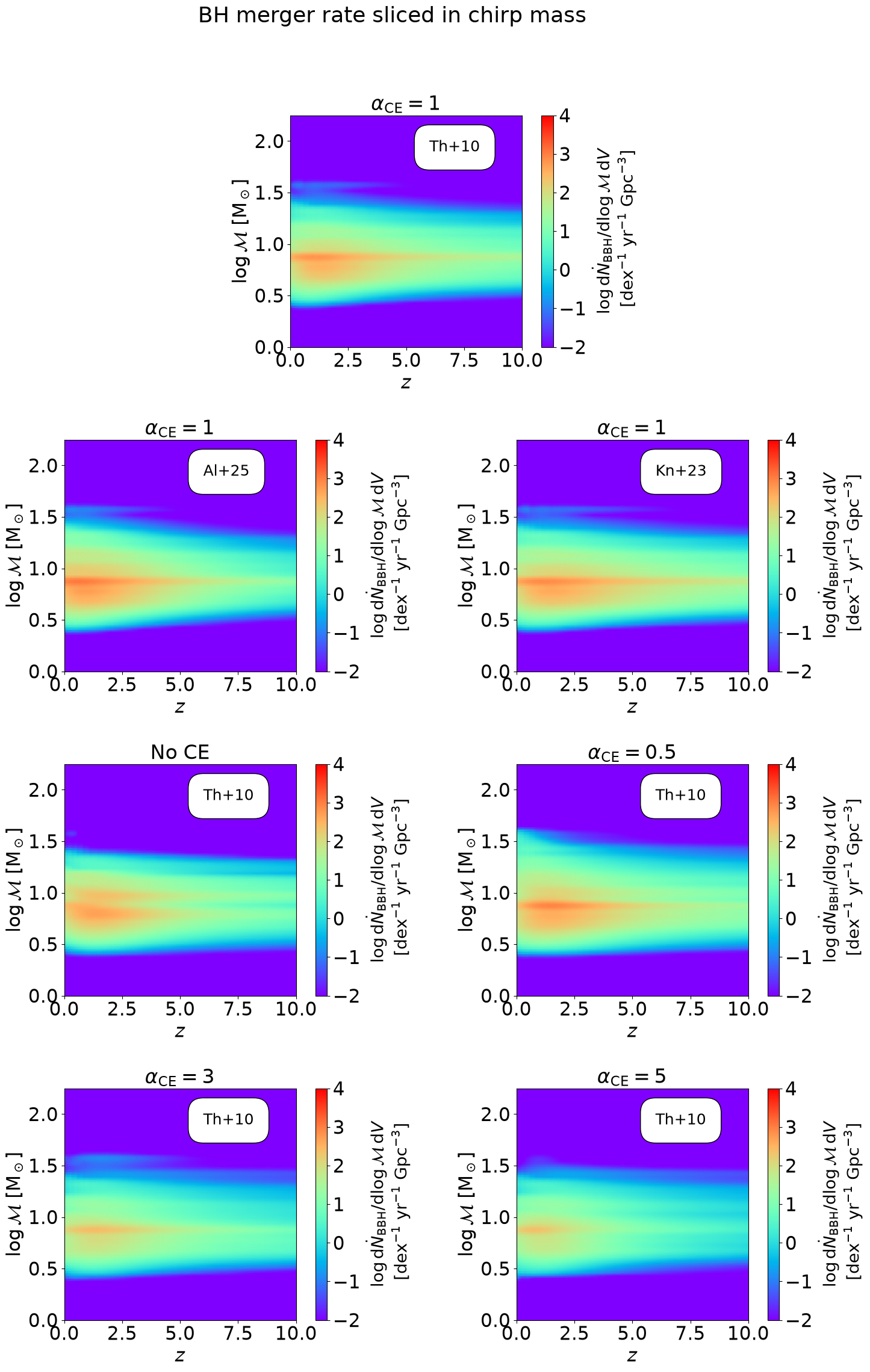}
\caption{Cosmic merger rate density of BBHs sliced in chirp mass and redshift. Panels as in Figure \ref{fig|BBHMR_2D_Z}.}\label{fig|BBHMR_2D_Mchirp}
\end{figure}

\clearpage

\bibliographystyle{JHEP}
\bibliography{ms}

\end{document}